\documentclass[preprint]{aastex63}

\usepackage{xspace}
\usepackage{enumitem}
\usepackage[utf8]{inputenc}
\usepackage{newunicodechar,graphicx}
\usepackage{amsmath}
\usepackage{lineno}
\usepackage{tabularx}
\usepackage{booktabs}
\usepackage{upgreek}

\DeclareRobustCommand{\okina}{%
  \raisebox{\dimexpr\fontcharht\font`A-\height}{%
    \scalebox{0.8}{`}%
  }%
}
\newunicodechar{ʻ}{\okina}

\newcommand{\dd}[1]{\mathrm{d}#1}

\newcommand{\kepler}{\textit{Kepler}\xspace}

\newcommand{\mesa}{\textsf{MESA}\xspace}
\newcommand{\yrec}{\textsf{YREC}\xspace}

\newcommand{\rocrit}{${\rm Ro}_{\rm crit}$\xspace}
\newcommand{\teff}{T$_{\rm eff}$\xspace}
\newcommand{\yini}{$Y_{\rm init}$\xspace}

\shorttitle{Weakened Magnetic Braking}
\shortauthors{Saunders et al.}

\graphicspath{{./}{figures/}}

\begin{document}

\title{Stellar Cruise Control: Weakened Magnetic Braking Leads to Sustained Rapid Rotation of Old Stars}

\author[0000-0003-2657-3889]{Nicholas Saunders}
\altaffiliation{NSF Graduate Research Fellow, nksaun@hawaii.edu}
\affiliation{Institute for Astronomy, University of Hawaiʻi at M\=anoa, 2680 Woodlawn Drive, Honolulu, HI 96822, USA}

\author[0000-0002-4284-8638]{Jennifer L. van Saders}
\affiliation{Institute for Astronomy, University of Hawaiʻi at M\=anoa, 2680 Woodlawn Drive, Honolulu, HI 96822, USA}

\author[0000-0001-8355-8082]{Alexander J. Lyttle}
\affiliation{School of Physics and Astronomy, University of Birmingham, Birmingham, B15 2TT, UK}

\author[0000-0003-4034-0416]{Travis S.~Metcalfe}
\affiliation{White Dwarf Research Corporation, 9020 Brumm Trail, Golden, CO 80403, USA}

\author{Tanda Li}
\affiliation{School of Physics and Astronomy, University of Birmingham, Birmingham, B15 2TT, UK}

\author{Guy R. Davies}
\affiliation{School of Physics and Astronomy, University of Birmingham, Birmingham, B15 2TT, UK}

\author[0000-0002-0468-4775]{Oliver J. Hall}
\affiliation{European Space Agency (ESA), European Space Research and Technology Centre (ESTEC), Keplerlaan 1, 2201 AZ Noordwijk, The Netherlands}

\author[0000-0002-4773-1017]{Warrick H. Ball}
\affiliation{School of Physics and Astronomy, University of Birmingham, Birmingham, B15 2TT, UK}
\affiliation{Advanced Research Computing, University of Birmingham, Edgbaston, Birmingham, B15 2TT, UK}

\author[0000-0002-2522-8605]{Richard Townsend}
\affiliation{Department of Astronomy, University of Wisconsin-Madison, 2535 Sterling Hall, 475 N. Charter Street, Madison, WI 53706, USA}
\affiliation{Kavli Institute for Theoretical Physics, University of California, Santa Barbara, CA 93106, USA}

\author{Orlagh Creevey}
\affiliation{Université Côte d’Azur, Observatoire de la Côte d’Azur, CNRS, Laboratoire Lagrange, Bd de l’Observatoire, CS 34229, 06304 Nice Cedex 4, France}

\author{Curt Dodds}
\affiliation{Institute for Astronomy, University of Hawaiʻi at M\=anoa, 2680 Woodlawn Drive, Honolulu, HI 96822, USA}

\begin{abstract}

   Despite a growing sample of precisely measured stellar rotation periods and ages, the strength of magnetic braking and the degree of departure from standard (Skumanich-like) spindown have remained persistent questions, particularly for stars more evolved than the Sun. Rotation periods can be measured for stars older than the Sun by leveraging asteroseismology, enabling models to be tested against a larger sample of old field stars. Because asteroseismic measurements of rotation do not depend on starspot modulation, they avoid potential biases introduced by the need for a stellar dynamo to drive starspot production. Using a neural network trained on a grid of stellar evolution models and a hierarchical model-fitting approach, we constrain the onset of weakened magnetic braking. We find that a sample of stars with asteroseismically-measured rotation periods and ages is consistent with models that depart from standard spindown prior to reaching the evolutionary stage of the Sun. We test our approach using neural networks trained on model grids produced by separate stellar evolution codes with differing physical assumptions and find that the choices of grid physics can influence the inferred properties of the braking law. We identify the normalized critical Rossby number ${\rm Ro}_{\rm crit}/{\rm Ro}_\odot = 0.91\pm0.03$ as the threshold for the departure from standard rotational evolution. This suggests that weakened magnetic braking poses challenges to gyrochronology for roughly half of the main sequence lifetime of sun-like stars.

\end{abstract}

\section{Introduction} \label{sec:intro}


Over their main sequence lifetimes, low-mass stars gradually lose angular momentum and slow their rotation due to magnetic braking \citep{weber,skumanich1972}. This angular momentum loss results from the interaction between a star's dynamo-generated field and stellar winds \citep{parker1958,kawaler1988,barnes2007}. The method of leveraging stellar rotation periods to estimate age, called \textit{gyrochronology} \citep{barnes2010,epstein2013}, can provide constraints on age with $\sim$10\% precision for sun-like stars in some age ranges \citep{meibom2015}. Numerous studies have provided prescriptions for angular momentum loss \citep{kawaler1988,krishnamurthi1997,sills2000,barnes2010,denissenkov2010,reiners2012,epstein2013,gallet2013,gallet2015,matt2015,vansaders2016}, which can be empirically calibrated to observations. The relationship between rotation period and age has been well characterized for young and intermediate-age clusters \citep{barnes2007,barnes2010,mamajek2008a,meibom2011,gallet2015,meibom2015,angus2019,dungee2022}, where both properties can be constrained with adequate precision. 

In essentially all of these calibrators, rotation rates are measured by observing spot modulation due to dark starspots rotating in and out of view. The high photometric precision of the \kepler Space Telescope \citep{borucki2010}, and the subsequent \textit{K2} mission \citep{howellK2MissionCharacterization2014}, enabled predictions for magnetic braking to be tested on a wealth of open clusters and associations (see \citealt{codyCatalog29Open2018}) as well as a population of older field stars \citep{mcquillan2014a,santos2021a}. 

In addition to starspot modulation used to detect rotation, brightness modulations due to stellar oscillations are measurable in the high-precision, long-baseline \kepler time series photometry \citep{huber2011}. Asteroseismology---the study of these oscillations---provides valuable information about the internal structure and evolution of stars. Specifically, stellar rotation rates can be measured from the mode frequencies \citep{nielsen2015,davies2015,hall2021} and ages can be inferred by comparisons with stellar models \citep{metcalfe2014a,metcalfe2016,silvaaguirre2015,creevey2017}.

When the ages of older, sun-like field stars were asteroseismically measured with \kepler data, they were found to maintain surprisingly rapid rotation late into their main sequence lifetimes \citep{angus2015}. To explain this sustained rapid rotation, it was proposed that stars diverge from the ``standard spindown" model and enter a phase of ``weakened magnetic braking" (WMB; \citealt{vansaders2016,vansaders2019}). When stellar rotation was measured using asteroseismology rather than spot modulation, the observed rotation periods were consistently faster than predicted by the standard spindown model and evidence for WMB strengthened \citep{hall2021}. Asteroseismology measures internal rotation rates in the stellar envelope, making it insensitive to surface differential rotation \citep{nielsen2015} and stellar inclination \citep{davies2015}; additionally, asteroseismology can measure rotation rates for stars with weak surface magnetic activity and therefore undetectable spot modulation signals \citep{chaplin2011}. These features allow asteroseismic rotation periods to avoid potential biases present in measurements from spot detection.

Careful analysis of pileups in the temperature-period distribution of sun-like stars also supported the WMB model. Studies of rotation rates in the \kepler field identified an upper envelope in stellar mass versus rotation period that matched a gyrochrone at $\sim$4 Gyr \citep{matt2015}. An upper edge to the distribution could be caused by either a magnetic transition or detection bias in spot modulation \citep{vansaders2019}. Forward modeling of the \kepler field predicted a pileup of rotation periods in the weakened braking scenario that was not seen in the data, but \cite{vansaders2019} argued that errors in the measured effective temperatures were obscuring the feature. With refined measurements of stellar effective temperature, the predicted pileup in the temperature-period distribution was identified \citep{david2022a}.

A study of sun-like stars with projected rotation periods measured from spectroscopic line broadening found them to be inconsistent with the Skumanich relation beyond $\sim$2 Gyr \citep{santos2016}, supporting a departure from standard spindown. This sample was later revisited \citep{lorenzo-oliveira2019}, and the analysis suggested that the smooth rotational evolution scenario was favored, and if weakened braking takes place, it occurs at later times ($\gtrsim5.3$ Gyr). However, these measurements faced biases introduced by an uncertain distribution of inclinations, which can inflate rotation periods measured spectroscopically. 

The physical mechanism that would lead to WMB remains uncertain, though some have proposed that a transition in the complexity of the magnetic field could reduce magnetic braking efficiency \citep{reville2015,garraffo2016,vansaders2016,metcalfe2016,metcalfe2019}. Because the transition may to be rooted in the strength and morphology of the magnetic field, it is challenging to test with surface rotation rates measured through spot modulation, which require active stellar dynamos to drive starspot production \citep{matt2015,reinhold2020}. 

To effectively use gyrochronology to estimate stellar ages, it is essential to understand when the transition to weakened braking occurs. Previous studies have provided estimates for the onset of WMB \citep{vansaders2016,vansaders2019,david2022a}, but fully hierarchical modeling for the braking law has not been previously performed. As the departure from standard spindown depends on the dimensionless Rossby number and is predicted to be shared between all stars \citep{vansaders2016}, the problem is inherently hierarchical. Here, we provide new constraints on the evolutionary phase at which stars undergo weakened braking. We build on previous efforts (e.g. \citealt{hall2021}) by modeling the rotational evolution of each star individually.

We apply a Hierarchical Bayesian Model (HBM) to constrain the population-level parameters for a WMB model. The use of an HBM has been shown to increase the precision of inferred stellar properties for high-dimensional models \citep{lyttle2021}. Here, we model the weakened braking parameters as global properties shared by all stars, while simultaneously fitting individual stellar properties. We test the results of our fit using multiple model grids, and compare the performance of a WMB model to standard spindown. By comparing results between multiple model grids, we provide the first constraints on biases introduced by the choices of grid physics when modeling stellar rotational evolution. We find that weakened braking likely occurs before stars reach the evolutionary phase of the Sun.

\section{Data} \label{sec:data}

We fit our rotational model to open clusters, the Sun, and \kepler field stars with asteroseismic measurements to ensure that we capture the early rotational evolution prior to the onset of weakened braking in addition to the behavior on the latter half of the main sequence. The seismic sample that best probes braking generally lies within $0.2$ M$_\odot$ of the Sun and covers a wide range of ages. Stars hotter than 6250 K ($\sim$1.2 M$_\odot$) lack deep convective envelopes on the main sequence, and do not undergo significant magnetic braking, and the seismic signals of stars cooler than 5000 K ($\sim$0.8 M$_\odot$) have low pulsation amplitudes and are challenging to measure. We describe our calibrator sources in the following section.

\subsection{Open Clusters} \label{sec:clusters}

\begin{figure*}
    \centering
    \gridline{\fig{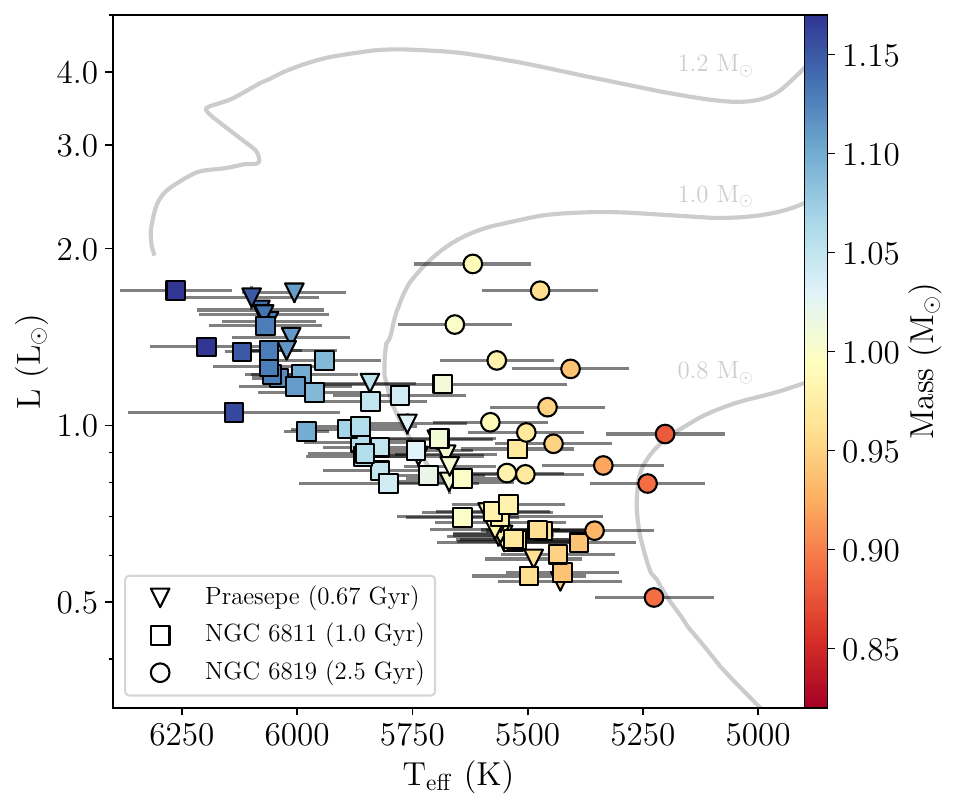}{.33\textwidth}{(a)}
              \fig{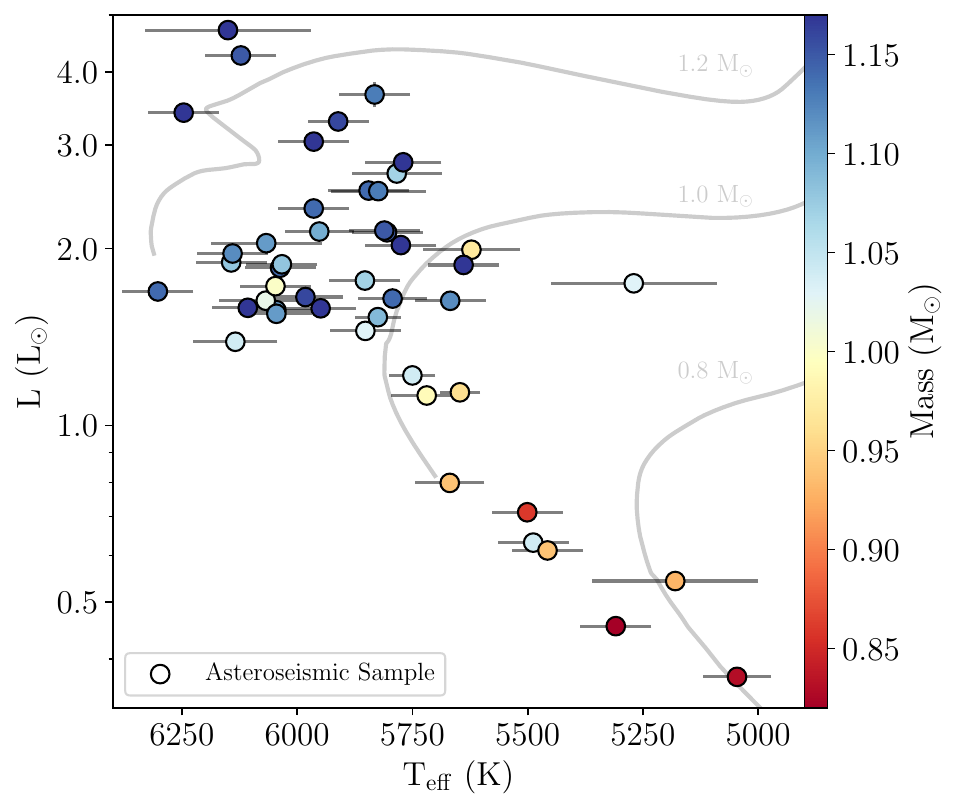}{.33\textwidth}{(b)}
              \fig{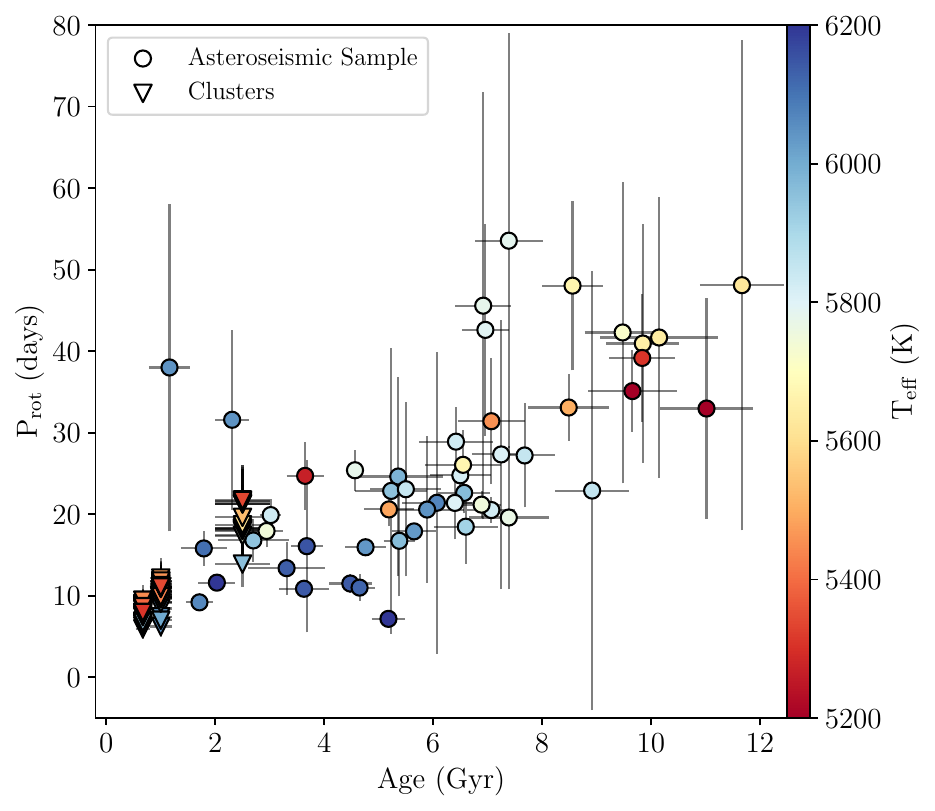}{.33\textwidth}{(c)}}
    \caption{\textbf{(a)} Hertzsprung-Russell diagram showing our sample of calibrators in open clusters. Model tracks generated by our emulator are shown as gray lines. \textbf{(b)} Hertzsprung-Russell diagram of our asteroseismic sample from Hall21. We derived the stellar properties shown here with asteroseismic modeling. \textbf{(c)} Observed rotation period plotted as a function of stellar age. We color points by their effective temperature. Asteroseismic stars are shown as circles and open cluster members are marked by triangles.}
    \label{fig:sample}
\end{figure*}

We included stars from the following open clusters: 23 stars in Praesepe ($0.67\pm0.134$ Gyr; [Fe/H] $=0.15\pm0.1$ dex; \citealt{rebull2017}), 45 stars in NGC 6811 ($1.0\pm0.2$ Gyr; [Fe/H] $=0.0\pm0.04$ dex; \citealt{meibom2011,curtis2019}), and 17 stars in NGC 6819 ($2.5\pm0.5$ Gyr; [Fe/H] $=0.10\pm0.03$ dex; \citealt{meibom2015}). We select stars within the \teff range of our asteroseismic sample (5200 K $\leq$ \teff $\leq$ 6200 K), using values for \teff reported in \cite{curtis2020}. Ages and metallicities were taken from the corresponding cluster reference, and were used to define priors in our fitting. The Hertzsprung-Russell diagram positions of the open cluster members can be seen in panel (a) of Figure \ref{fig:sample}.

\subsection{Asteroseismic Sample} \label{sec:astero_sample}

We also included a sample of \kepler field stars with asteroseismically-measured rotation rates and ages from \citet[hereafter Hall21]{hall2021}. Rotation rates for main sequence stars can be challenging to measure with starspot modulation, particularly for older and less active stars, due to long rotation periods and diminished stellar activity. However, the rotational splitting of asteroseismic oscillation frequencies can be observed for stars in the end stages of the main sequence, and provides invaluable benchmarks for WMB. 

\cite{hall2021} used asteroseismic mode splitting to measure rotation periods for 91 \kepler dwarfs. We augmented the Hall21 sample with two additional stars with asteroseismic rotation measurements in the wide binary system HD 176465 (KIC 10124866; \citealt{white2017}). The A and B components of this system are sometimes referred to by their nicknames \textit{Luke} and \textit{Leia}, respectively. The rotation periods reported in \cite{white2017} were derived by fitting asteroseismic mode splitting, following the same approach as Hall21.

We performed asteroseismic modeling for Luke \& Leia and 47 stars from the Hall21 sample that fall within our desired mass range using version 2.0 of the Asteroseismic Modeling Portal\footnote{\href{https://urldefense.com/v3/__https://github.com/travismetcalfe/amp2__;!!PvDODwlR4mBZyAb0!Qrc9ClC0O4rSXcFMupyp4SE3ReG5o2aoKnmlHiTZ41SD3_s1krek1ZK2AQ26fDTnp39wsm9Zp2qvAuzN_fY$}{github.com/travismetcalfe/amp2}} \citep[AMP;][]{Metcalfe2009, Woitaszek2009,Metcalfe2023}. This optimization method couples a parallel genetic algorithm \citep{Metcalfe2003} with MESA stellar evolution models \citep{paxton2019} and the GYRE pulsation code \citep{townsend2013} to determine the stellar properties that most closely reproduce the observed oscillation frequencies and spectroscopic constraints for each star. The choices of input physics are nearly all the default choices in MESA release 12778, and the models include gravitational settling of helium and heavy elements \citep{Thoul1994} as well as the two-term correction for surface effects proposed by \cite{Ball2014}. The resulting asteroseismic sample is shown 
in panel (b) of Figure \ref{fig:sample}, while the stellar properties and rotation periods can be found in Table \ref{tab:astero}, which includes maximum-likelihood estimates of the age, mass, composition, and mixing-length from our AMP modeling. 

With masses derived from asteroseismic modeling, we made mass cuts ($0.8$ M$_\odot \leq$ M $\leq 1.2$ M$_\odot$) to ensure our sample would fall within the bounds of our model grids. Previous studies have indicated that rotation periods in field stars $<7$ days are likely due to non-eclipsing short-period binaries \citep{simonian2019,simonian2020}, and we therefore remove three stars (KIC 6603624, KIC 8760414, KIC 8938364) from the sample that showed rotation $<7$ days at ages $>8$ Gyr that we suspect are inconsistent with single star evolution. Panel (c) of Figure \ref{fig:sample} shows the rotation periods and ages for our full sample of open clusters and asteroseismic field stars.

\begin{table}
\scriptsize
    \begin{tabular}{l  r  r  r  r  r  r  r}
    \toprule
        KIC & Age (Gyr) & $P_{\rm rot}$ (days) & $M$ ($M_\odot$) & T$_{\rm eff}$ (K) & [Fe/H] (dex) & Y$_{\rm init}$ & $\alpha_{\rm MLT}$ \\
        \midrule 
        10644253 & $1.16 \pm 0.38$ & $38.01 \pm 20.11$ & $1.11 \pm 0.05$ & $6045 \pm 77$ & $0.06 \pm 0.10$ & $0.29 \pm 0.03$ & $1.66 \pm 0.10$ \\
        8379927 & $1.71 \pm 0.25$ & $9.20 \pm 0.24$ & $1.11 \pm 0.04$ & $6067 \pm 120$ & $-0.10 \pm 0.15$ & $0.26 \pm 0.02$ & $1.84 \pm 0.15$ \\
        3735871 & $1.79 \pm 0.42$ & $15.81 \pm 2.14$ & $1.17 \pm 0.04$ & $6107 \pm 77$ & $-0.04 \pm 0.10$ & $0.23 \pm 0.02$ & $1.74 \pm 0.09$ \\
        9139151 & $2.03 \pm 0.34$ & $11.61 \pm 0.94$ & $1.14 \pm 0.02$ & $6302 \pm 77$ & $0.10 \pm 0.10$ & $0.26 \pm 0.02$ & $1.70 \pm 0.09$ \\
        3427720 & $2.31 \pm 0.31$ & $31.59 \pm 11.03$ & $1.11 \pm 0.04$ & $6045 \pm 77$ & $-0.06 \pm 0.10$ & $0.27 \pm 0.02$ & $1.72 \pm 0.08$ \\
        10079226 & $2.70 \pm 0.65$ & $16.79 \pm 2.62$ & $1.20 \pm 0.04$ & $5949 \pm 77$ & $0.11 \pm 0.10$ & $0.23 \pm 0.02$ & $1.72 \pm 0.11$ \\
        10124866B & $2.94 \pm 0.29$ & $17.95 \pm 2.00$ & $0.97 \pm 0.03$ & $5745 \pm 94$ & $-0.30 \pm 0.06$ & $0.24 \pm 0.01$ & $1.84 \pm 0.13$ \\
        10124866A & $3.02 \pm 0.20$ & $19.90 \pm 2.00$ & $0.95 \pm 0.02$ & $5831 \pm 93$ & $-0.30 \pm 0.06$ & $0.25 \pm 0.01$ & $1.62 \pm 0.10$ \\
        4141376 & $3.31 \pm 0.70$ & $13.39 \pm 3.24$ & $1.04 \pm 0.04$ & $6134 \pm 91$ & $-0.24 \pm 0.10$ & $0.25 \pm 0.02$ & $1.82 \pm 0.12$ \\
        8394589 & $3.63 \pm 0.46$ & $10.86 \pm 0.58$ & $1.08 \pm 0.02$ & $6143 \pm 77$ & $-0.29 \pm 0.10$ & $0.26 \pm 0.01$ & $1.78 \pm 0.10$ \\
        9025370 & $3.65 \pm 0.34$ & $24.71 \pm 4.17$ & $1.03 \pm 0.03$ & $5270 \pm 180$ & $-0.12 \pm 0.18$ & $0.25 \pm 0.01$ & $2.58 \pm 0.18$ \\
        10730618 & $3.68 \pm 0.29$ & $16.09 \pm 10.58$ & $1.19 \pm 0.03$ & $6150 \pm 180$ & $-0.11 \pm 0.18$ & $0.27 \pm 0.01$ & $1.76 \pm 0.15$ \\
        10963065 & $4.48 \pm 0.39$ & $11.51 \pm 1.14$ & $1.12 \pm 0.03$ & $6140 \pm 77$ & $-0.19 \pm 0.10$ & $0.24 \pm 0.02$ & $1.74 \pm 0.08$ \\
        8228742 & $4.65 \pm 0.29$ & $11.00 \pm 1.64$ & $1.15 \pm 0.06$ & $6122 \pm 77$ & $-0.08 \pm 0.10$ & $0.25 \pm 0.01$ & $1.40 \pm 0.10$ \\
        6106415 & $4.76 \pm 0.37$ & $15.95 \pm 0.74$ & $1.14 \pm 0.02$ & $6037 \pm 77$ & $-0.04 \pm 0.10$ & $0.23 \pm 0.01$ & $1.72 \pm 0.06$ \\
        8694723 & $5.18 \pm 0.31$ & $7.17 \pm 0.72$ & $1.17 \pm 0.03$ & $6246 \pm 77$ & $-0.42 \pm 0.10$ & $0.22 \pm 0.01$ & $1.76 \pm 0.12$ \\
        8006161 & $5.19 \pm 0.46$ & $20.60 \pm 2.04$ & $1.04 \pm 0.02$ & $5488 \pm 77$ & $0.34 \pm 0.10$ & $0.25 \pm 0.01$ & $1.98 \pm 0.09$ \\
        5094751 & $5.23 \pm 0.66$ & $22.86 \pm 17.53$ & $1.10 \pm 0.03$ & $5952 \pm 75$ & $-0.08 \pm 0.10$ & $0.28 \pm 0.02$ & $1.66 \pm 0.11$ \\
        11133306 & $5.36 \pm 0.82$ & $24.63 \pm 12.17$ & $1.16 \pm 0.04$ & $5982 \pm 82$ & $-0.02 \pm 0.10$ & $0.23 \pm 0.02$ & $1.80 \pm 0.12$ \\
        12258514 & $5.38 \pm 0.29$ & $16.75 \pm 6.83$ & $1.17 \pm 0.02$ & $5964 \pm 77$ & $0.00 \pm 0.10$ & $0.23 \pm 0.01$ & $1.44 \pm 0.06$ \\
        4914423 & $5.50 \pm 0.65$ & $23.07 \pm 10.69$ & $1.14 \pm 0.03$ & $5845 \pm 88$ & $0.07 \pm 0.11$ & $0.27 \pm 0.02$ & $1.68 \pm 0.16$ \\
        6116048 & $5.65 \pm 0.40$ & $17.90 \pm 1.02$ & $1.08 \pm 0.02$ & $6033 \pm 77$ & $-0.23 \pm 0.10$ & $0.25 \pm 0.01$ & $1.76 \pm 0.07$ \\
        9410862 & $5.89 \pm 0.67$ & $20.58 \pm 8.99$ & $1.00 \pm 0.03$ & $6047 \pm 77$ & $-0.31 \pm 0.10$ & $0.26 \pm 0.02$ & $1.80 \pm 0.12$ \\
        7106245 & $6.07 \pm 0.64$ & $21.41 \pm 18.50$ & $1.02 \pm 0.02$ & $6068 \pm 102$ & $-0.99 \pm 0.19$ & $0.22 \pm 0.02$ & $1.76 \pm 0.13$ \\
        4914923 & $6.40 \pm 0.51$ & $21.39 \pm 4.47$ & $1.15 \pm 0.03$ & $5805 \pm 77$ & $0.08 \pm 0.10$ & $0.24 \pm 0.02$ & $1.68 \pm 0.08$ \\
        6933899 & $6.42 \pm 0.68$ & $28.91 \pm 4.27$ & $1.13 \pm 0.03$ & $5832 \pm 77$ & $-0.01 \pm 0.10$ & $0.27 \pm 0.02$ & $1.68 \pm 0.11$ \\
        6521045 & $6.50 \pm 0.56$ & $24.78 \pm 1.94$ & $1.13 \pm 0.02$ & $5824 \pm 103$ & $0.02 \pm 0.10$ & $0.26 \pm 0.02$ & $1.68 \pm 0.09$ \\
        3544595 & $6.55 \pm 0.70$ & $26.06 \pm 4.29$ & $0.94 \pm 0.03$ & $5669 \pm 75$ & $-0.18 \pm 0.10$ & $0.25 \pm 0.02$ & $1.88 \pm 0.11$ \\
        10516096 & $6.57 \pm 0.48$ & $22.62 \pm 2.52$ & $1.14 \pm 0.02$ & $5964 \pm 77$ & $-0.11 \pm 0.10$ & $0.24 \pm 0.01$ & $1.72 \pm 0.07$ \\
        11401755 & $6.60 \pm 0.59$ & $18.48 \pm 4.52$ & $1.16 \pm 0.02$ & $5911 \pm 66$ & $-0.20 \pm 0.06$ & $0.22 \pm 0.02$ & $1.72 \pm 0.13$ \\
        12069449 & $6.89 \pm 0.35$ & $21.18 \pm 1.64$ & $1.04 \pm 0.01$ & $5750 \pm 50$ & $0.05 \pm 0.02$ & $0.26 \pm 0.01$ & $1.84 \pm 0.05$ \\
        7296438 & $6.92 \pm 0.51$ & $45.59 \pm 26.21$ & $1.18 \pm 0.03$ & $5775 \pm 77$ & $0.19 \pm 0.10$ & $0.24 \pm 0.01$ & $1.74 \pm 0.06$ \\
        11295426 & $6.96 \pm 0.43$ & $42.61 \pm 13.02$ & $1.14 \pm 0.02$ & $5793 \pm 74$ & $0.12 \pm 0.07$ & $0.22 \pm 0.01$ & $1.76 \pm 0.05$ \\
        12069424 & $7.07 \pm 0.44$ & $20.52 \pm 1.54$ & $1.09 \pm 0.02$ & $5825 \pm 50$ & $0.10 \pm 0.03$ & $0.25 \pm 0.01$ & $1.76 \pm 0.05$ \\
        9955598 & $7.07 \pm 0.62$ & $31.41 \pm 7.72$ & $0.94 \pm 0.03$ & $5457 \pm 77$ & $0.05 \pm 0.10$ & $0.25 \pm 0.02$ & $1.92 \pm 0.12$ \\
        7680114 & $7.25 \pm 0.54$ & $27.34 \pm 16.52$ & $1.15 \pm 0.02$ & $5811 \pm 77$ & $0.05 \pm 0.10$ & $0.24 \pm 0.02$ & $1.74 \pm 0.07$ \\
        10586004 & $7.39 \pm 0.73$ & $19.60 \pm 8.81$ & $1.17 \pm 0.04$ & $5770 \pm 83$ & $0.29 \pm 0.10$ & $0.28 \pm 0.02$ & $2.16 \pm 0.20$ \\
        10514430 & $7.39 \pm 0.62$ & $53.56 \pm 25.42$ & $1.07 \pm 0.05$ & $5784 \pm 98$ & $-0.11 \pm 0.11$ & $0.27 \pm 0.03$ & $1.78 \pm 0.09$ \\
        9098294 & $7.68 \pm 0.55$ & $27.21 \pm 6.37$ & $1.03 \pm 0.02$ & $5852 \pm 77$ & $-0.18 \pm 0.10$ & $0.25 \pm 0.01$ & $1.86 \pm 0.08$ \\
        7871531 & $8.49 \pm 0.74$ & $33.09 \pm 4.10$ & $0.86 \pm 0.02$ & $5501 \pm 77$ & $-0.26 \pm 0.10$ & $0.28 \pm 0.01$ & $1.94 \pm 0.11$ \\
        3656476 & $8.56 \pm 0.56$ & $48.04 \pm 10.40$ & $1.12 \pm 0.02$ & $5668 \pm 77$ & $0.25 \pm 0.10$ & $0.26 \pm 0.01$ & $1.80 \pm 0.04$ \\
        5950854 & $8.92 \pm 0.68$ & $22.91 \pm 26.94$ & $1.07 \pm 0.04$ & $5853 \pm 77$ & $-0.23 \pm 0.10$ & $0.22 \pm 0.01$ & $1.92 \pm 0.12$ \\
        8424992 & $9.48 \pm 0.70$ & $42.30 \pm 18.51$ & $0.99 \pm 0.03$ & $5719 \pm 77$ & $-0.12 \pm 0.10$ & $0.23 \pm 0.02$ & $1.90 \pm 0.09$ \\
        11772920 & $9.66 \pm 0.81$ & $35.11 \pm 4.99$ & $0.93 \pm 0.05$ & $5180 \pm 180$ & $-0.09 \pm 0.18$ & $0.23 \pm 0.01$ & $2.16 \pm 0.28$ \\
        7970740 & $9.84 \pm 0.61$ & $39.17 \pm 7.85$ & $0.81 \pm 0.02$ & $5309 \pm 77$ & $-0.54 \pm 0.10$ & $0.27 \pm 0.02$ & $2.14 \pm 0.11$ \\
        11904151 & $9.85 \pm 0.67$ & $40.94 \pm 14.66$ & $0.96 \pm 0.02$ & $5647 \pm 44$ & $-0.15 \pm 0.10$ & $0.25 \pm 0.02$ & $1.84 \pm 0.07$ \\
        8349582 & $10.15 \pm 1.08$ & $41.68 \pm 17.26$ & $1.17 \pm 0.05$ & $5639 \pm 77$ & $0.30 \pm 0.10$ & $0.23 \pm 0.03$ & $1.86 \pm 0.11$ \\
        6278762 & $11.02 \pm 0.85$ & $32.97 \pm 13.53$ & $0.83 \pm 0.02$ & $5046 \pm 74$ & $-0.37 \pm 0.09$ & $0.22 \pm 0.02$ & $2.42 \pm 0.19$ \\
        4143755 & $11.67 \pm 0.77$ & $48.10 \pm 30.07$ & $0.97 \pm 0.04$ & $5622 \pm 106$ & $-0.40 \pm 0.11$ & $0.23 \pm 0.02$ & $1.78 \pm 0.13$ \\
        \bottomrule
    \end{tabular}
\caption{The sample of 49 asteroseismic \kepler field stars used in our fit. The rotation periods for 10124866A \& B are from \cite{white2017}, all other rotation periods taken from \cite{hall2021}. T$_{\rm eff}$ and [Fe/H] are adopted from the LEGACY \citep{lund2017,silvaaguirre2015} and KAGES \citep{aguirre2015,davies2016} catalogs (see \citealt{hall2021}). Other stellar properties were derived for this work using asteroseismic mode fitting.}
\label{tab:astero}
\end{table}

\section{Methods} \label{sec:methods}

We produced model grids for rotational evolution using two stellar evolution codes---Modules for Experiments in Stellar Astrophysics (\mesa; \citealt{paxton2010,paxton2013,paxton2015,paxton2018,paxton2019}) and Yale Rotating Stellar Evolution Code (\yrec; \citealt{pinsonneault1989,demarque2008a}). The ranges of stellar properties covered by our grid are detailed in Table \ref{tab:grid}, and we describe the model physics used to generate each grid in the following sections.

\begin{table}[ht!]
\footnotesize
\centering
    \begin{tabular}{l l  c  c}
    \toprule
        Parameter && \mesa Bounds & \yrec Bounds \\ \midrule
        Mass & $M$ ($M_\odot$) & [0.8, 1.2] & [0.8, 1.2] \\
        Mixing Length Parameter & $\alpha_{\rm MLT}$ & [1.4, 2.0] & [1.4, 2.0] \\
        Metallicity & [Fe/H] (dex) & [-0.3, 0.3] & [-0.3, 0.3] \\
        Initial Helium Abundance & \yini & [0.22, 0.28] & \textit{not varied} \\
        Braking Law Strength & $f_K$ & [4.0, 11.0] & [4.0, 11.0] \\
        Critical Rossby Number & Ro$_{\rm crit}$ & [1.0, 4.5] & [1.0, 4.5]\\
        \bottomrule
    \end{tabular}
    \caption{Parameter boundaries of the \mesa and \yrec grids.}
    \label{tab:grid}
\end{table}

\subsection{\mesa Model Grid} \label{sec:mesa}

We construct our \mesa grid with identical input physics to the models used for asteroseismic inference (described in \S \ref{sec:astero_sample}) in order to avoid biases introduced by the modeling (see \citealt{tayar2020}). Our models used initial elemental abundances from \cite{grevesse1998} and an atmospheric temperature structure following an Eddington $T(\uptau)$ relation with fixed opacity. We smoothly ramp diffusion from fully modeled at M $\leq$ 1.1 M$_\odot$ to no diffusion at M $\geq$ 1.2 M$_\odot$. We do not include core or envelope overshoot. We varied the mass $M$, metallicity [Fe/H], initial Helium abundance \yini, and mixing length parameter $\alpha_{\rm MLT}$. 

We calculated rotational evolution histories (as described in \S \ref{sec:braking_model}) for each combination of stellar properties and appended them to our grid. By default, \mesa models do not output the necessary stellar parameters to perform rotational evolution, and it was necessary to adapt the outputs included in the grid. The additional parameters we include for each star were the total moment of inertia $I_{\rm tot}$, the moment of inertia of the convective envelope $I_{\rm env}$, the photospheric pressure $P_{\rm phot}$, and the convective overturn timescale $\uptau_{\rm cz}$. We define $\uptau_{\rm cz}$ as
\[
\uptau_{\rm cz} = \frac{H_P}{v_{\rm conv}}
\]
where $H_P$ is the pressure scale height at the convective zone boundary and $v_{\rm conv}$ is the convective velocity one pressure scale height above the base of the convective zone. 

Stellar interiors in \mesa models are divided into shells and the parameters are evaluated at a finite number of points. We identified the precise location of the base of the convective zone as a function of the star's mass fraction using the Schwarzschild criterion, and then interpolated between the values calculated at each shell boundary to more precisely identify the values of our desired parameters at each time step. 

\subsection{\yrec Model Grid} \label{sec:yrec}

We construct our \yrec grid following the settings laid out in \cite{vansaders2013} and \cite{metcalfe2020}. We use the mixing length theory of convection \citep{vitense1953,cox1968} with the 2006 OPAL equation of state \citep{rogers1996,rogers2002}. Abundances were taken from \cite{grevesse1998} and opacities from the Opacity Project \citep{mendoza2007}. We define atmosphere and boundary conditions from \cite{kurucz1997}. Nuclear reaction rates were drawn from \cite{adelberger2011}. \yini was fixed to a linear Helium-enrichment law anchored to the Sun with a slope of $\left(\frac{\dd Y}{\dd Z}\right)_\odot=1.296$ (see \S \ref{sec:grid_bias}). We varied the same parameters as we did for the \mesa grid, with the exception of \yini.

As with the \mesa grid, we trace additional parameters to evaluate the angular momentum loss law. For each model at each timestep, we calculate the moment of inertia of both the star and its convective envelope, the photospheric pressure, and the convective overturn timescale. 

\subsection{Magnetic Braking Model} \label{sec:braking_model}

Prescriptions for magnetic braking often incorporate the dimensionless Rossby Number (Ro), defined as the ratio between the rotation period, $P$, and convective overturn timescale within the stellar envelope, $\uptau_{\rm cz}$, as a means to estimate magnetism across stars of different masses. We use the Rossby number in our rotation model due to its utility as a tracer for both the mass and composition dependence of spindown and magnetic field strength. We invoke a Rossby threshold, \rocrit, beyond which point stars depart from a simple power law spindown and conserve angular momentum \citep{vansaders2016}. We adopt the \cite{matt2012} modification to the \cite{kawaler1988} braking law. We assume, as in \cite{vansaders2013}, that the magnetic field strength $B$ scales as $B\propto$ P$_{\rm phot}^{1/2}$Ro$^{-1}$, where P$_{\rm phot}$ is the photospheric pressure, and that mass loss $\dot{M}$ scales as $\dot{M}\propto L_X \propto L_{\rm bol}$Ro$^{-2}$, where $L_X$ is the x-ray luminosity and $L_{\rm bol}$ is the bolometric luminosity.

Our full model for rotational evolution is described by
\begin{equation*}
\frac{\dd J}{\dd t} = 
   \begin{cases} 
      f_K K_M \omega \left( \frac{\omega_{\text{sat}}}{\omega_\odot} \right)^2, & \omega_{\rm sat} \leq \omega \frac{\uptau_{\rm cz}}{\uptau_{{\rm cz}, \odot}}, {\rm Ro} \leq {\rm Ro}_{\rm crit} \\
      f_K K_M \omega \left( \frac{\omega\uptau_{\rm cz}}{\omega_\odot\uptau_{{\rm cz},\odot}} \right)^2, & \omega_{\rm sat} > \omega \frac{\uptau_{\rm cz}}{\uptau_{{\rm cz}, \odot}}, {\rm Ro} \leq {\rm Ro}_{\rm crit} \\
      0, & {\rm Ro} > {\rm Ro}_{\rm crit}
   \end{cases}
\end{equation*}
where Ro is defined as
\begin{equation*}
    {\rm Ro} = \frac{P}{\uptau_{\rm cz}},
\end{equation*}
$f_K$ is the scaling factor for the strength of angular momentum loss during classical spindown, $\omega_{\rm sat}$ is the threshold at which angular momentum loss saturates for young stars, and with
\begin{equation*}
    \frac{K_M}{K_{M,\odot}} = c(\omega)\left(\frac{R}{R_\odot}\right)^{3.1}\left(\frac{M}{M_\odot}\right)^{-0.22}\left(\frac{L}{L_\odot}\right)^{0.56}\left(\frac{P_{\rm phot}}{P_{\rm phot},\odot}\right)^{0.44}.
\end{equation*}
The term $c(\omega)$ is the centrifugal correction from \cite{matt2012}, and we assume $c(\omega)=1$, which is appropriate for slowly rotating stars.

To calculate the rotation histories for our grid, we take the outputs of non-rotating \mesa and \yrec models, and compute rotation periods with the \textsf{rotevol} code \citep{vansaders2013,somers2017}. We focus only on $f_K$ and \rocrit as they will be the most dominant parameters of a WMB law for the stars in our sample, which are old enough to have converged onto tight rotation sequences \citep{epstein2013,gallet2015}. We assume a disk locking period of 8.13 days and disk lifetime of 0.28 Myr, setting the initial rotation rates of our models \citep{vansaders2013}. We fix $\omega_{\rm sat}$ to 3.863 $\times 10^{-5}$ rad/s. Each of these parameters will be important at early ($<100$ Myr) times, but will have negligible effects by the time stars reach the ages in our sample. We assume solid body rotation in our models, since the epoch of radial differential rotation in this mass range is again limited to young stars \citep{denissenkov2010,gallet2015,spada2020}.

\subsection{Model Grid Emulator} \label{sec:ANN}

With rotationally evolved model grids, we construct an emulator for rapid stellar evolution modeling. The general approach to this type of optimization problem is simple interpolation between tracks in a high-dimensional model grid (e.g. \citealt{berger2020}). However, due to the size of the grid, number of parameters (4-5 per star and cluster, with 2 additional global braking law parameters), and large sample of potential targets, this approach becomes computationally expensive, particularly in the application of Bayesian inference through sampling the model. We therefore opt to train an artificial neural network (ANN) to map the stellar parameters of the grid to observable parameters of stars in our sample. 

We define our \mesa ANN with seven input parameters and four output parameters. Our inputs represent fundamental stellar properties: age, mass, metallicity, initial Helium abundance, mixing length parameter, braking law strength, and critical Rossby number. The ANN outputs are observable quantities: effective temperature, radius, surface metallicity, and rotation period. The \yrec ANN has the above input parameters with \yini excluded, and identical output parameters. The remainder of this section describes the training and characterization of the \mesa ANN. The process for training the \yrec ANN is identical, and we compare the results when using different grids in \S \ref{sec:grid_bias}.

Our model structure results in a neural network that acts as a stellar evolution emulator. Given some set of input stellar properties, the model will output the corresponding observable quantities. Because the emulation is rapid, the model can also be used to calculate likelihoods to infer input parameters---given some set of observed properties, we can sample prior distributions for the underlying stellar properties and retrieve posterior distributions, providing estimates for these values with uncertainties.

We construct an ANN with 6 hidden layers comprised of 128 neurons each (following the tuning process of \citealt{lyttle2021}). Each hidden layer used an Exponential Linear Unit (ELU) activation function. Using \textsf{TensorFlow} \citep{abadi2016}, we trained the model on an NVidia Tesla V100 graphics processing unit (GPU) for 10,000 epochs using an \textit{Adam} optimizer \citep{kingma2017} with a learning rate of $10^{-5}$. We trained the ANN in $\sim$8,000 batches of $\sim$16,000 points. The full model architecture is detailed in Appendix \ref{sec:nn_structure}.

Prior to training the ANN, we remove the pre-main sequence from the tracks in our grid, defined as the threshold at which the luminosity from nuclear burning exceeds 99\% of the total stellar luminosity. We allow the tracks to begin evolving across the subgiant branch, as our sample includes stars at or approaching this evolutionary stage, but remove tracks that exceed a rotation period of 150 days. 

\begin{figure*}[ht!]
    \centering
    \includegraphics[width=.9\textwidth]{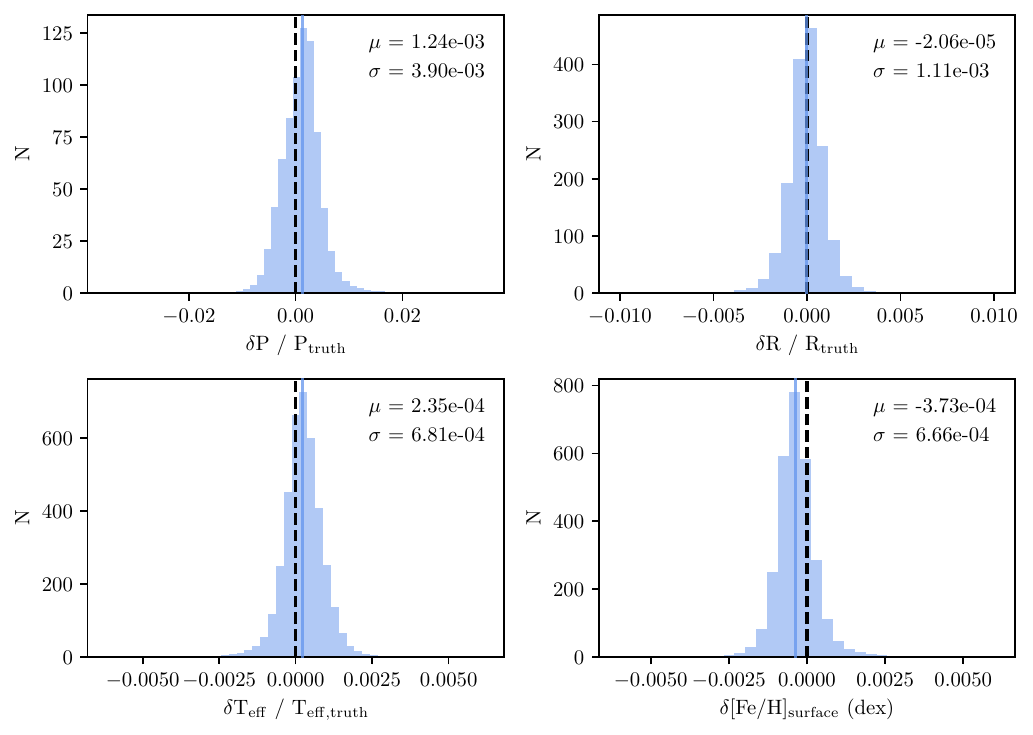}
    \caption{Uncertainty introduced by the \mesa ANN emulator. The histograms for P, R, and \teff show the (predicted$-$truth)/truth value for our training set, and the bottom right panel shows predicted$-$truth for the surface metallicity to account for points where [Fe/H]$_{\rm surface,truth}\approx0$. The median $\mu$ and standard deviation $\sigma$ of these distributions are shown in the top right corner for each parameter, and $\mu$ is marked by the solid vertical line. The error incurred by the ANN is negligible compared to the uncertainty on the observed values.}
    \label{fig:nn_bias}
\end{figure*}

In order to ensure that the mapping performed by the neural network does not introduce significant uncertainty to the inferred parameters, we divide the grid data into a training set and a validation set. The training set is composed of 80\% of the models in the grid, drawn at random, and is used to generate the connections between the input model parameters and observed stellar properties. The remaining 20\% of the grid is then used as a validation set to predict the observed parameters based on the provided input parameters, allowing us to characterize the neural network's ability to successfully predict well understood values. When compared to the measurement uncertainties associated with these parameters, the error introduced by the ANN is negligible, with typical fractional uncertainties of $\sim$$10^{-3}$ in the recovery of our validation set (see Figure \ref{fig:nn_bias}). We also find negligible systematic offset for parameters in our validation set, indicating that the ANN is not introducing significant bias. 

\subsection{Statistical Modelling} \label{sec:stats_model}

In order to efficiently optimize the braking law model parameters, we construct a hierarchical Bayesian model (HBM). The application of a similar HBM for constraining the distribution of \yini and $\alpha_{\rm MLT}$ has been demonstrated by \cite{lyttle2021}. We begin the construction of our model with Bayes' theorem---the posterior probability of our model parameters $\boldsymbol{\theta}_i$ given some set of observed data $\mathbf{d}_i$ is
\begin{equation*}
    p(\boldsymbol{\theta}_i|\mathbf{d}_i) \propto p(\boldsymbol{\theta}_i)p(\mathbf{d}_i|\boldsymbol{\theta}_i)
\end{equation*}
where $p(\boldsymbol{\theta}_i)$ is the prior on the model parameter $\boldsymbol{\theta}_i$ (for $i$ parameters) and $p(\mathbf{d}_i|\boldsymbol{\theta}_i)$ is the likelihood of the data given the model. We use our trained ANN to sample the prior distribution $p(\boldsymbol{\theta}_i)$ for each parameter and evaluate an instance of the model $\boldsymbol{\mu}_i=\boldsymbol{\lambda}_i(\boldsymbol{\theta}_i)$, where $\boldsymbol{\lambda}_i$ represents the ANN model. From this, we can represent the likelihood of each observation $\mathbf{d}_i$ with uncertainty $\boldsymbol{\sigma}_i$ given the model evaluation $\boldsymbol{\mu}_i$ as the normal distribution 
\begin{equation*}
    p(\mathbf{d}_i|\boldsymbol{\theta}_i) = \prod_{n=1}^{N} \frac{1}{\sigma_{n,i}\sqrt{2\pi}} \exp{\left[ - \frac{(d_{n,i}-\mu_{n,i})^2}{2\sigma_{n,i}^2} \right]}
\end{equation*}
given $N$ observed variables. 

The hierarchical structure of our model allows us to prescribe various levels of pooling to different parameters. The WMB model parameters $f_K$ and \rocrit, for example, are assumed to be the same for all stars in our sample. For the ANNs trained on both the the \mesa and \yrec grids, we define the prior for \rocrit as
\begin{equation*}
    {\rm Ro_{crit}} \sim \mathcal{U}(1.0, 4.5)
\end{equation*}
and the prior for $f_K$ as
\begin{equation*}
    f_K \sim \mathcal{U}(4.0, 11.0)
\end{equation*}
where $\theta \sim X$ represents a parameter $\theta$ being randomly drawn from a distribution $X$, and $\mathcal{U}(a, b)$ is a uniform distribution bounded between $a$ and $b$. The values of \rocrit and $f_K$ drawn from these uniform distributions are used to calculate the full set of model evaluations $\boldsymbol{\mu}_i$ for that step. The bounds for \rocrit and $f_K$ were centered near the solar Rossby number derived for our grids (for \mesa: Ro$_\odot \approx 2.05$, $f_{K,\odot}\approx5.89$; for \yrec: Ro$_\odot \approx 2.33$, $f_{K,\odot}\approx7.52$). 

Other parameters are assumed to be unique to each star. For the \yrec ANN, we constrain the mass, metallicity, mixing length parameter, and age. We constrain the same parameters for the \mesa ANN with the addition of the initial Helium abundance. The prior distributions for these parameters are defined as truncated normal distributions, given by
\begin{equation*}
    p(\boldsymbol{\theta}) \sim \mathcal{N}_{[a,b]}(\mu, \sigma)
\end{equation*}
where $\mathcal{N}$ is the normal distribution, $a$ and $b$ are the lower and upper bounds, respectively, $\mu$ is the median and $\sigma$ is the standard deviation. Here, $\mu$ and $\sigma$ are taken from the observational constraints on the parameters and their uncertainties. For stars in clusters, we define a prior centered on the value reported in the corresponding reference (see \S \ref{sec:clusters}) with a width set to the measurement uncertainty for age, metallicity, mixing length parameter, and rotation period (with the inclusion of \yini for the \mesa grid). For the masses of cluster stars, we use a homology scaling relationship with \teff and set a broad prior ($\sigma_M=$0.25 M$_\odot$), and for the mixing length parameter and initial Helium abundance we use uniform priors. For asteroseismic stars in our sample, all of the above properties are constrained by the asteroseismic fitting, and we use this asteroseismic value and its uncertainty as the center and width of the prior distributions, respectively. Our truncated distributions for all stars are bounded by the grid limits described in Table \ref{tab:grid}.

Finally, we include a third class of prior distributions in our model which are shared by some stars but not all. Each star within the same cluster is assumed to have the same age, metallicity, and initial Helium abundance, while these parameters should be fully independent for each target in the asteroseismic sample and for the Sun. These prior distributions share the same truncated normal form as the independent parameters, but can be selectively applied to specific subsets of the data.

With our priors and likelihoods defined, we sampled the model parameters. The ANN is compatible with automatic differentiation, allowing us to utilize No-U-Turn Sampling (NUTS; \citealt{nuts}). We constructed a probabilistic model with PyMC3 \citep{salvatier2016}, then calculated the maximum a posteriori estimate as our starting point and sampled 4 chains for 5,000 draws with 1,000 tuning steps. We sampled chains long enough to ensure that the Gelman-Rubin $\hat{R}$ statistic \citep{gelman1992} was lower than 1.01 for all parameters indicating model convergence. The residuals from our fit, as well as an example of our model fit to the Sun, are shown in Appendix \ref{sec:validation}.

\section{Results} \label{sec:results}

We optimize the parameters of our model under two different assumptions---standard spindown and WMB. In the standard spindown framework, we assume stars follow a Skumanich-like angular momentum loss law, where $\dot{J} \propto \omega^3$ at late times. Under the WMB assumption, stars lose angular momentum to magnetized stellar winds with the same relation as the standard spindown law until they reach a critical Rossby number \rocrit, at which point angular momentum is conserved. We use the \mesa ANN as our primary emulator as its grid physics match the models used in the asteroseismic parameter estimates. In the standard spindown case, we only optimize for $f_K$, and retrieve a constraint of $f_K=6.11 \pm 0.73$. For the WMB model, we report $f_K=5.46 \pm 0.51$ and \rocrit/Ro$_\odot$ $=0.91 \pm 0.03$.

\begin{figure*}[ht!]
    \centering
    \includegraphics[width=.9\textwidth]{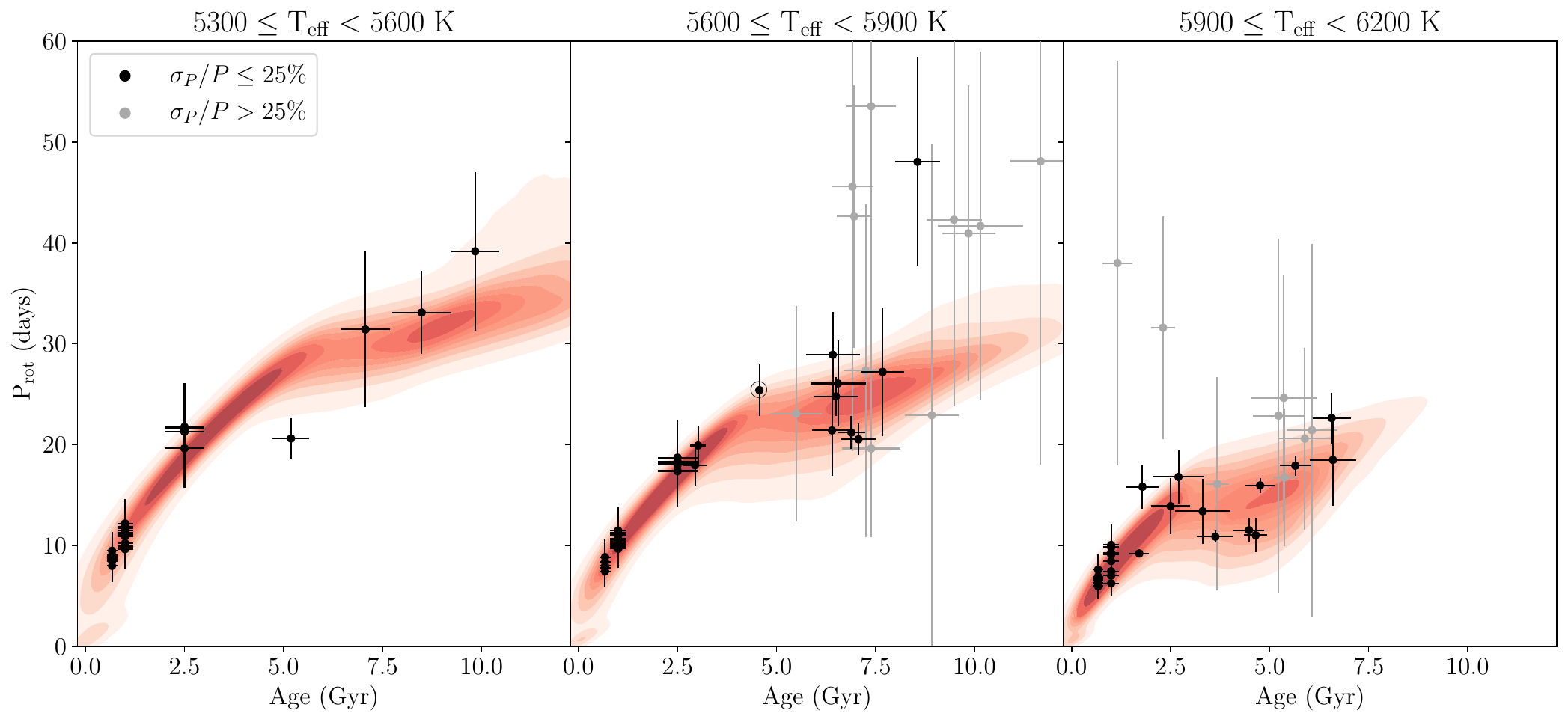}
    \caption{Stellar rotation period versus age, shown in three bins each spanning 300 K in \teff. Asteroseismic measurements and cluster stars are shown by points---black points represent rotation periods with fractional uncertainties $\sigma_P/P\leq 25\%$ and gray points show $\sigma_P/P> 25\%$. The Sun is marked by the $\odot$ symbol. Red contours represent the distribution of rotation periods within a given \teff bin predicted by our \mesa emulator model, produced from a sample of one million emulated stars with stellar properties randomly drawn from uniform distributions bounded by our sample, and $f_K$ and \rocrit fixed to the median values of the posterior distributions.}
    \label{fig:per_age}
\end{figure*}

Figure \ref{fig:per_age} shows the distribution of rotation periods predicted by our WMB model. We have divided the sample into equal-size bins in \teff because temperature captures the effects of both a star's mass and metallicity on its rotational evolution. The red shaded regions show the density of stars drawn from a simulated population of 1,000,000 stars under the best-fit WMB assumptions, generated with stellar properties drawn from uniform distributions for each parameter bounded by the edges of our sample using our \mesa emulator. The width of the distribution is caused by the range of masses, metallicities, Helium abundances, and mixing length parameters within each \teff bin. Stars in clusters can be seen as groups with discrete, well-constrained ages below 2.5 Gyr, and are valuable calibrators for the early angular momentum loss $\dot{J}$. In our model, this early $\dot{J}$ is captured by the braking law strength parameter, $f_K$. Stars in our asteroseismic sample span a wide range of ages, particularly on the second half of the main sequence, and provide the constraint on \rocrit. 

\begin{figure*}[ht!]
    \centering
    \includegraphics[width=.9\textwidth]{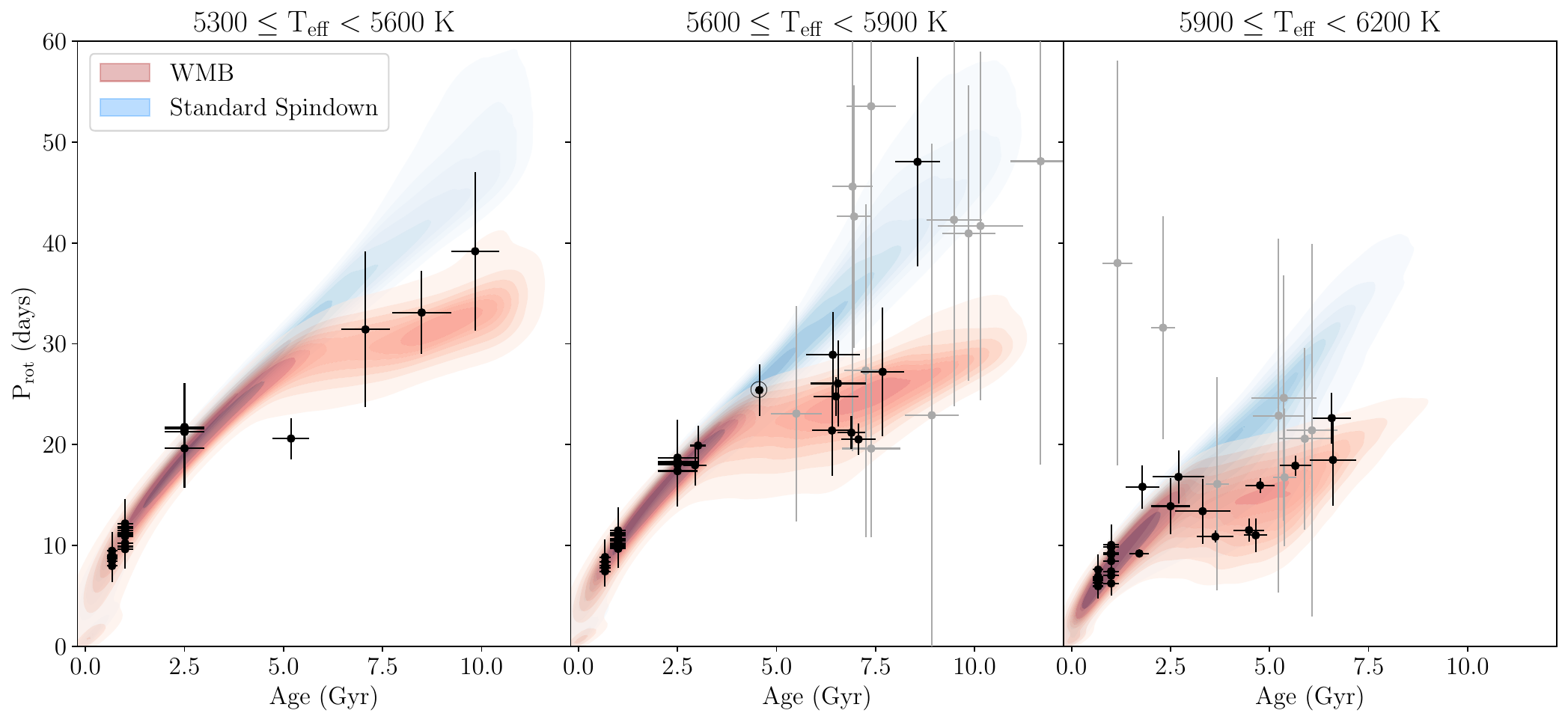}
    \caption{Same as Figure \ref{fig:per_age}, with the additional comparison to the standard spindown model. The contours represent the distribution of predicted rotation periods within a given \teff bin, with red showing our WMB model and blue showing a standard Skumanich-like spindown model, both generated with our \mesa emulator. The value of $f_K$ for the standard spindown model is the median value from the posterior of a fit to our full sample with no \rocrit constraint.}
    \label{fig:pvt_v_standard}
\end{figure*}

In Figure \ref{fig:pvt_v_standard}, we show the comparison between the rotation periods predicted by both the standard spindown and WMB models (in blue and red, respectively). Each shaded region represents the density of points in a population of 100,000 simulated stars from our \mesa emulator. The standard spindown model was fit to the full sample, without altering angular momentum loss beyond a Rossby threshold. The models produce similar constraints on $f_K$, as the early rate of $\dot{J}$ is well-constrained by the clusters in both models. At older ages, the standard spindown model significantly overpredicts the rotation periods of stars in our asteroseismic sample.

\begin{figure*}[ht!]
    \centering
    \includegraphics[width=.5\textwidth]{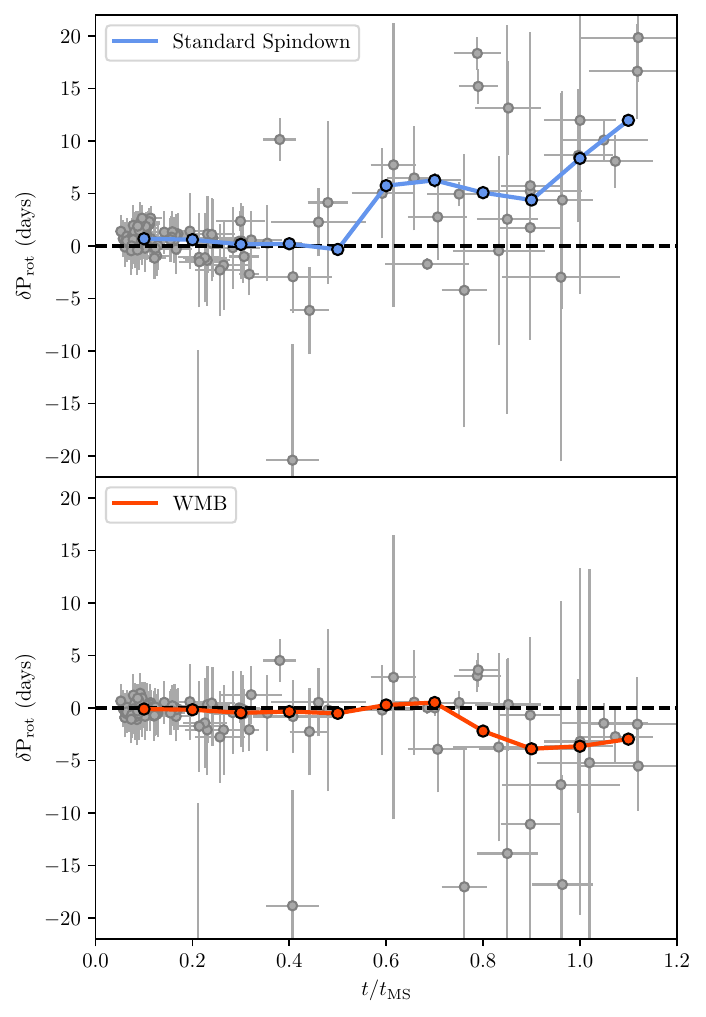}
    \caption{Difference between predicted and observed rotation periods for all stars in our sample (shown as gray points) as a function of fraction of main sequence lifetime $t/t_{\rm MS}$. The blue and red points represent the uncertainty-weighted median of $\delta P$ within a 0.2 $t/t_{\rm MS}$ bin for the standard and WMB models, respectively. Main sequence lifetime was estimated by identifying the age of core-H exhaustion in \mesa models generated for each star. Roughly halfway through the main sequence lifetime, the standard spindown model begins significantly over-predicting rotation periods. The WMB model is consistent with the observed distribution until near the end of the main sequence, at which point it underestimates rotation periods.}
    \label{fig:compare_tms}
\end{figure*}

The WMB model results in a smaller average deviation from the observed rotation periods. Figure \ref{fig:compare_tms} shows the the difference between predicted and observed rotation periods for our sample. The colored points show the uncertainty-weighted median within a 0.2 $t/t_{\rm MS}$ bin. On average, the standard spindown model overpredicts rotation periods by 0.72 days for the full sample and 6.00 days for stars beyond the first half of the main sequence ($t/t_{\rm MS} \geq 0.5$). Conversely, WMB underpredicts rotation periods by 0.31 days for the full sample and 3.18 days for stars past $0.5t/t_{\rm MS}$. Isolating only the asteroseismic sample (at all ages), standard spindown overpredicts $P_{\rm rot}$ by 4.66 days on average, and WMB underpredicts by 2.02 days. The corresponding fractional deviations for the asteroseismic sample are $+17.73\%$ for standard spindown and $-9.09\%$ for WMB. We perform a reduced chi-squared test to determine the goodness-of-fit for our models, and we find $\chi^2_{\rm \nu, WMB}=1.07$ and $\chi^2_{\rm \nu, standard}=14.02$. Because $\chi^2_{\rm \nu, WMB} \ll \chi^2_{\rm \nu, standard}$, we conclude that the WMB model provides a better fit to the data.

Figure \ref{fig:compare_tms} shows the difference between predicted and observed rotation periods as a function of fraction of main sequence lifetime. For the first half of the main sequence, the standard spindown and WMB models both describe the observed rotation periods well. However, at roughly halfway though the main sequence ($0.5t/t_{\rm MS}$), the standard spindown model deviates from the observed distribution and begins overpredicting rotation periods. Both models are consistent with the cluster data, which follow a tight spindown sequence that is nearly identical for the two models (see Figure \ref{fig:pvt_v_standard}). 

\section{Discussion} \label{sec:discussion}

We have provided refined probabilistic estimates for the onset of WMB, described by the parameter \rocrit. Our model indicates that stars enter a phase of weakened braking before reaching the Rossby number of the Sun (\rocrit$ = 0.91\pm0.03$ Ro$_\odot$). This result supports constraints by \cite{david2022a}, which found a sub-solar \rocrit when examining the pileup in the temperature-period distribution of \kepler stars. \cite{vansaders2016,vansaders2019} found that a critical Rossby number of \rocrit $\approx$ Ro$_\odot$ provided the best fit to the observed rotation periods, which agrees with our results within 2$\sigma$.

The new constraints on weakened braking parameters provided here can be used as guidelines for where gyrochronology is likely to be accurate. Beyond \rocrit, rotation evolves only slowly with the changing moment of inertia, and stars can be observed with the same rotation period for Gyr timescales, challenging any gyrochronological estimate. We show that gyrochonological ages should be precise until $\sim$Ro$_\odot$, corresponding to an age of $\sim$4 Gyr for sun-like stars. After the onset of WMB, age estimates should have significantly larger uncertainties due to the slowly evolving rotation on the second half of the main sequence. 

\subsection{WMB Model Performance} \label{sec:wmb_performance}

Towards the end of the main sequence, our model for weakened braking begins to underestimate rotation periods. This likely reflects our overly simple implementation of the transition from standard to weakened braking. The immediate shutdown of angular momentum loss beyond \rocrit is the simplest model which introduces the fewest new parameters. Given the limited sample of reliable calibrators spanning a wide range of \teff near the onset of WMB, any parameterization of a possible gradual transition, or a transition that does not completely shut down magnetic braking, is not well constrained. As more seismic constraints are placed on the ages and rotation periods near \rocrit, additional parameters that lead to a gradual transition, or $\dot{J}\neq0$ beyond the transition, can be tested. 

The deviation between the WMB model and observed rotation periods could additionally be partially explained by small deviations in inferred model ages. At the end of a star's main sequence lifetime, even in the WMB framework when angular momentum is conserved, the rotation period increases steeply due to the changing stellar moment of inertia as the star's radius expands. Models for rotation increase on short time spans in parallel vertical tracks in rotation-age space as stars traverse the subgiant branch, with small separations between stars of different \teff. Improved asteroseismic modeling, or a larger sample of stars with asteroseismic parameter constraints, could better distinguish between these effects at the end of the main sequence.

\subsection{Assessing the Asteroseismic Constraint} \label{sec:seis_constraint}

\begin{figure*}[ht!]
    \centering
    \includegraphics[width=.5\textwidth]{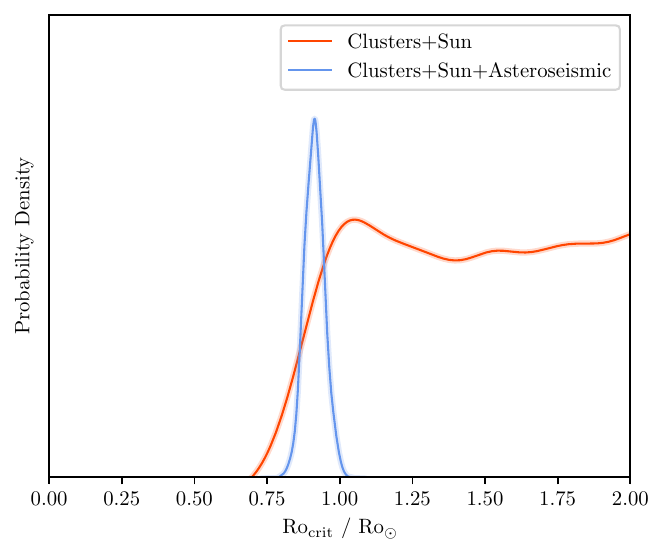}
    \caption{Comparison between posterior distributions for \rocrit from models fit to different subsets of the data. When fit to only clusters and the Sun (shown in red), \rocrit is unconstrained beyond the solar Rossby number. With the inclusion of the asteroseismic sample (shown in blue), \rocrit is tightly constrained just below the solar Rossby number. The y-axis has been arbitrarily scaled for clarity.}
    \label{fig:kde}
\end{figure*}

To illustrate the impact of the asteroseismic sample on our ability to constrain \rocrit, we fit our model to two subsets of the data: one comprised of only clusters and the Sun, capturing the early rotational evolution, and one that adds the asteroseismic stars. Figure \ref{fig:kde} shows a Kernel Density Estimate (KDE) of the sampled marginal posterior distributions for \rocrit when fit to each of these samples. When fit to only clusters and the Sun, \rocrit has little to no likelihood below the solar value, and is unconstrained beyond the solar value. This aligns with our expectations, as the young cluster sample has repeatedly been shown to follow standard braking \citep{barnes2007,barnes2010,mamajek2008a,gallet2015,meibom2011,meibom2015}. When the asteroseismic sample is included, the posterior becomes tightly constrained near the solar value. This exercise clearly demonstrates why the effects of WMB were not identified until a large enough sample of stars with precise rotation periods and ages spanning the main sequence were available. 

\subsection{Consistency with Solar Twins} \label{sec:solar_twins}

A recent study by \cite{lorenzo-oliveira2019} proposed tension between the weakened magnetic braking model and an observed population of ``solar twins." The stars in this sample have typical masses within $\pm 0.05$ M$_\odot$ of solar and metallicities with $\pm 0.04$ dex of solar. Rotation periods were not directly measured for the majority of stars in this sample, instead the projected rotational velocity $v\sin{i}$ of each star was estimated from spectral line broadening. This was converted to a projected rotation period, $P_{\rm rot} / \sin{i}$, using stellar properties derived from from \textit{Gaia} DR2 \citep{gaiacollaboration2018} and ground-based spectroscopic data.

\begin{figure*}[ht!]
    \centering
    \gridline{\fig{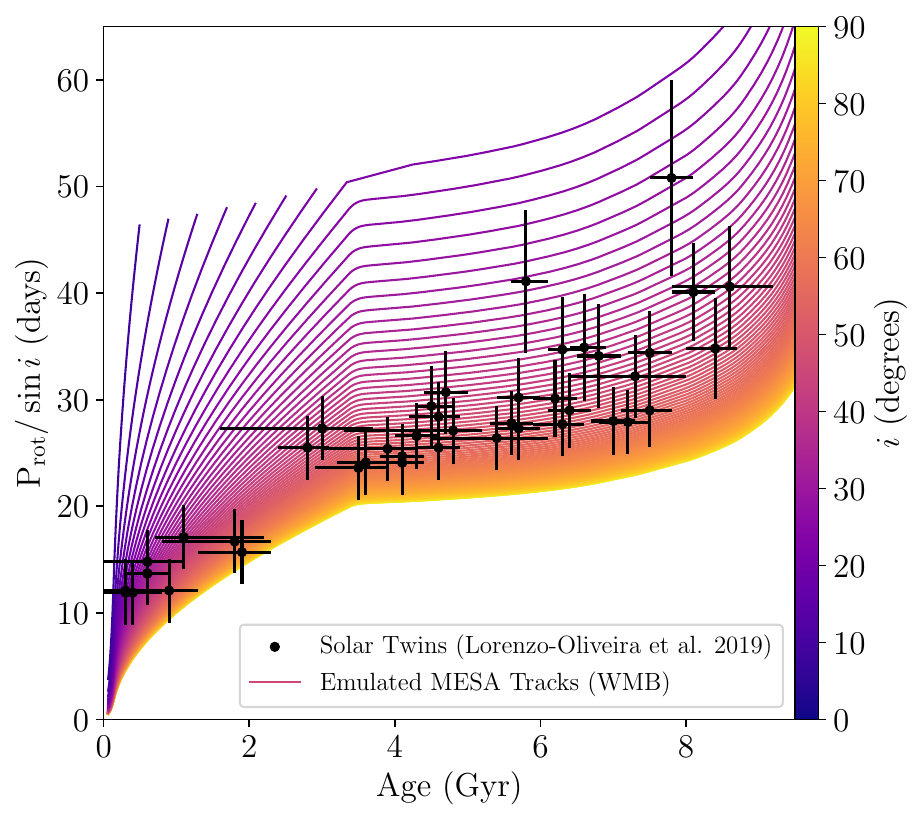}{.32\textwidth}{(a)}
              \fig{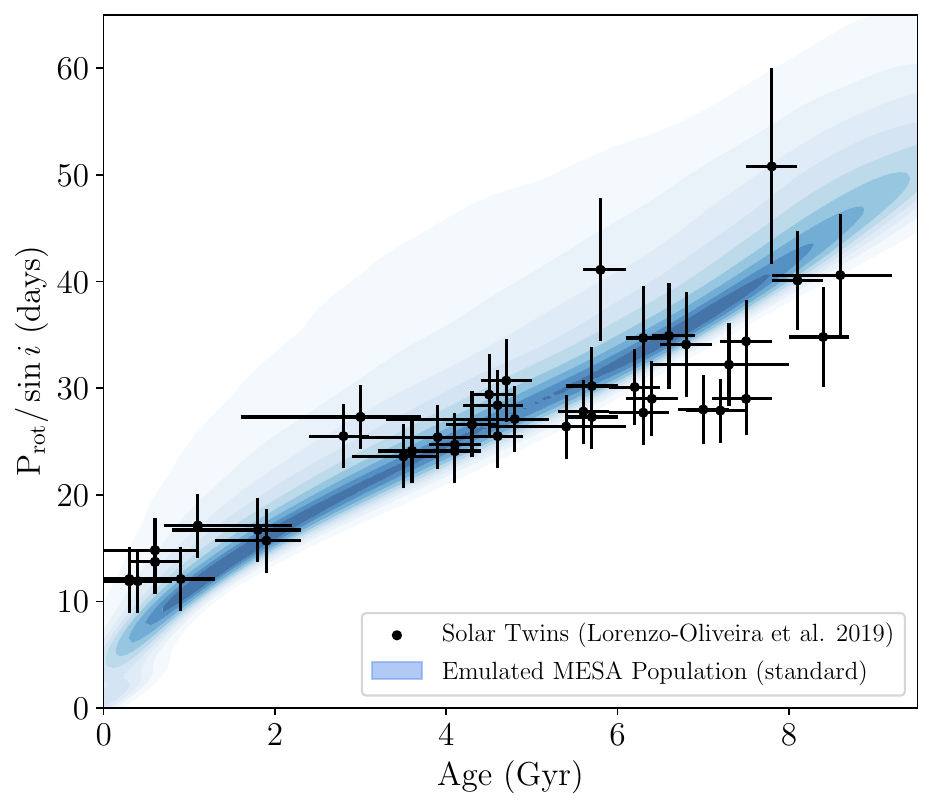}{.32\textwidth}{(b)}
              \fig{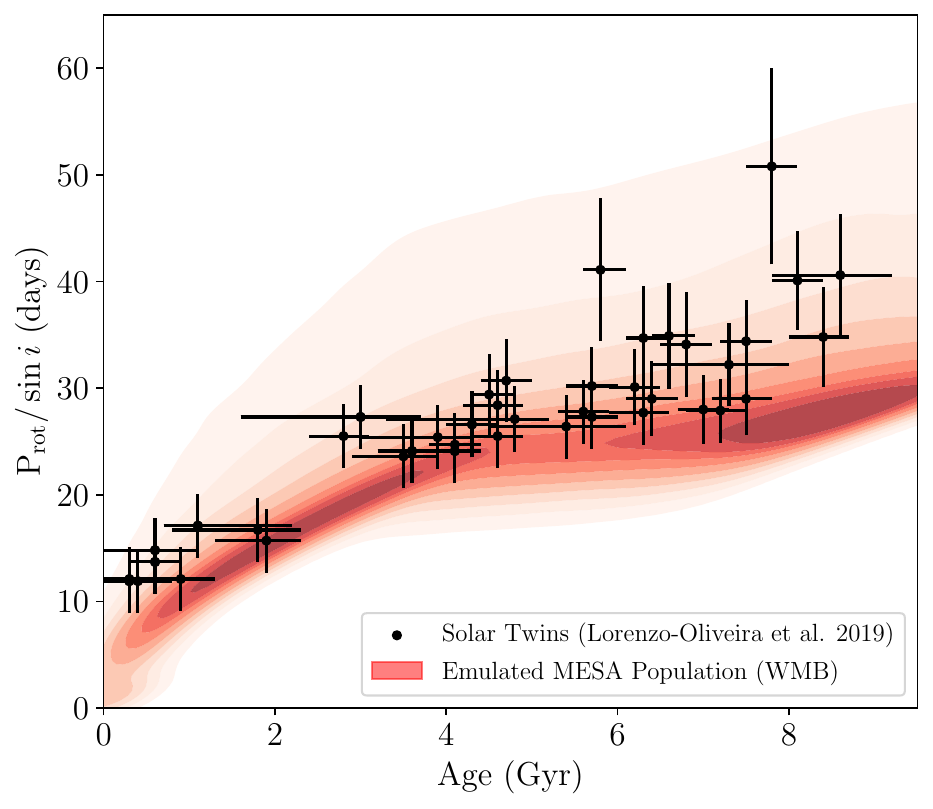}{.32\textwidth}{(c)}}
    \caption{\textbf{(a)} Projected rotation period, $P_{\rm rot} / \sin{i}$, of the solar twins sample versus age. The colored lines show tracks from our MESA emulator for a solar-calibrated model with a range of stellar inclinations, evolved under WMB assumptions. Models that are not observed edge on have their projected periods shifted to higher values. \textbf{(b)} The solar twins sample compared to a standard spindown model with a range of stellar inclinations. We generated a population of 1,000,000 stars with parameters drawn from uniform distributions within $\pm0.05$ M$_\odot$ of solar for M, $\pm0.04$ dex of solar for [Fe/H], and inclinations, $i$, drawn from a uniform distribution in $\cos{i}$. \yini and $\alpha_{\rm MLT}$ were drawn from uniform distributions covering our model grid. \textbf{(c)} Same as panel (b), but with the WMB model. $f_K$ and \rocrit were fixed to values fit to our full sample. The standard model overpredicts rotation periods of the solar twins sample beyond the age of the Sun, while they are consistent with WMB when accounting for inclinations.}
    \label{fig:solartwins}
\end{figure*}

If a system is observed directly edge on ($i=90^\circ$), the projected rotation period will match that measured from photometric spot modulation or asteroseismic mode splitting. The primary effect of rotation axis inclination away from $90^\circ$ is to shift the projected rotation period to a higher value (see panel (a) of Figure \ref{fig:solartwins}). \cite{lorenzo-oliveira2019} undergo a selection process of simulating projected rotation periods given some random orientation between 0 and $90^\circ$, comparing their measured population against these simulations, and reducing their sample to stars they found most likely to be seen edge on based on the agreement (see \S 2 of \citealt{lorenzo-oliveira2019} for a full description of their approach). As only a fraction of the observed sample is likely to be observed directly edge on, the fastest rotation periods in the solar twins sample represent a lower envelope to the true distribution of rotation periods of the sample. 

We test the standard spindown and WMB models against the solar twins sample, seen in panels (b) and (c) of Figure \ref{fig:solartwins}. We calculate $P_{\rm rot} / \sin{i}$ for our \mesa emulator model tracks, drawing inclinations randomly from a uniform distribution between 0 and 1 in $\cos{i}$. The stellar properties of our model grid were drawn from uniform distributions bounded by the parameter cuts described in \cite{lorenzo-oliveira2019}---mass and metallicity were bounded by $0.8$ M$_\odot \leq$ M $\leq 1.2$ M$_\odot$ and $-0.04 \leq$ [Fe/H] $\leq +0.04$, and unconstrained parameters were given broad uniform priors ($0.22 \leq$ \yini $\leq 0.28$, $1.4 \leq \alpha_{\rm MLT} \leq 2.0$). We note that fixing \yini and $\alpha_{\rm MLT}$ to solar-calibrated values has negligible impact on the model fit. We find that the standard spindown model overpredicts projected rotation periods beyond the age of the Sun. The WMB model predicts the observed population with minor deviations from entirely edge-on inclinations. We find that the WMB model reasonably reproduces the behavior observed in the solar twins, and does so better than the standard spindown model.

\subsection{Accounting for Grid Bias} \label{sec:grid_bias}

We test our model fit using neural networks trained on grids of models generated by two stellar evolution codes, \mesa and \yrec. This provides an opportunity to independently validate our results as well as test for any bias introduced by the choice of grid. To date, most investigations of WMB have used ages and rotational evolution that were inferred using reasonable, but different, underlying stellar evolution models. Our \mesa grid was constructed with input physics matching the asteroseismic modeling, avoiding the cross-grid bias when fitting the \mesa-trained neural network to the asteroseismic observations. While we have matched the physics in the seismic and rotational models, we have not performed the fits simultaneously, which we reserve for future work.

The primary difference between the construction of the grids was to vary Y$_{\rm init}$ as an additional dimension of the \mesa grid, while calculating it with a fixed He-enrichment law in the \yrec grid. We used a relation to compute the value of Y$_{\rm init}$ for a model in the \yrec grid given its metallicity [Fe/H] given by
\[
Y_{\rm init} = Y_P + \frac{(1-Y_P)\left(\frac{\dd Y}{\dd Z}\right)_\odot}{\left(\frac{\dd Y}{\dd Z}\right)_\odot + \left(\frac{Z}{X}\right)_\odot^{-1} 10^{-{\rm [Fe/H]}} + 1}
\]
where Y$_P$ is the primordial Helium abundance, the slope of the Helium enrichment law that matches the solar value is $\left(\frac{\dd Y}{\dd Z}\right)_\odot=1.296$, and the solar metal fraction is $\left(\frac{Z}{X}\right)_{\odot}=0.02289$ \citep{grevesse1998}.

The ANN for the \yrec grid was trained identically to the process for the \mesa grid described in \S \ref{sec:ANN}, and we constructed the probabilistic model following the process described in \S \ref{sec:stats_model}. For the \yrec ANN, the value of \yini fit by our asteroseismic modeling with \mesa was not used as a constraint on the model likelihood, while it was for the \mesa ANN. The choice to include \yini as a free parameter, as well as the differences between how different stellar evolution codes calculate quantities used in our modeling, have the potential to introduce systematic biases in the resulting model fits. Here, we compare between the results inferred by emulators trained on different model grids.

Most braking laws include a strong Ro dependence, and thus a dependence on the convective overturn timescale $\uptau_{\rm cz}$, and there is no single agreed upon means of calculating this value (see \citealt{kim1996}). Furthermore, changes in grid physics can result in different values of $\uptau_{\rm cz}$, even in solar-calibrated models. To account for this, we normalize \rocrit by a grid-dependent solar Rossby number Ro$_\odot$. To calculate Ro$_\odot$ for each grid, we produced solar-calibrated stellar evolution tracks and compute the Rossby number at the age of the Sun. For each model grid, we also compute the value of $f_K$ that reproduced solar rotation at solar age under the standard spindown assumption, and apply this as a normalization factor when comparing the inferred values of $f_K$ in our WMB models. We notate this solar-normalized braking law strength as $f_K'$. These normalization factors allow us to compare directly between the braking law parameters inferred from the ANN trained on each model grid.

\begin{figure*}[ht!]
    \centering
    \gridline{\fig{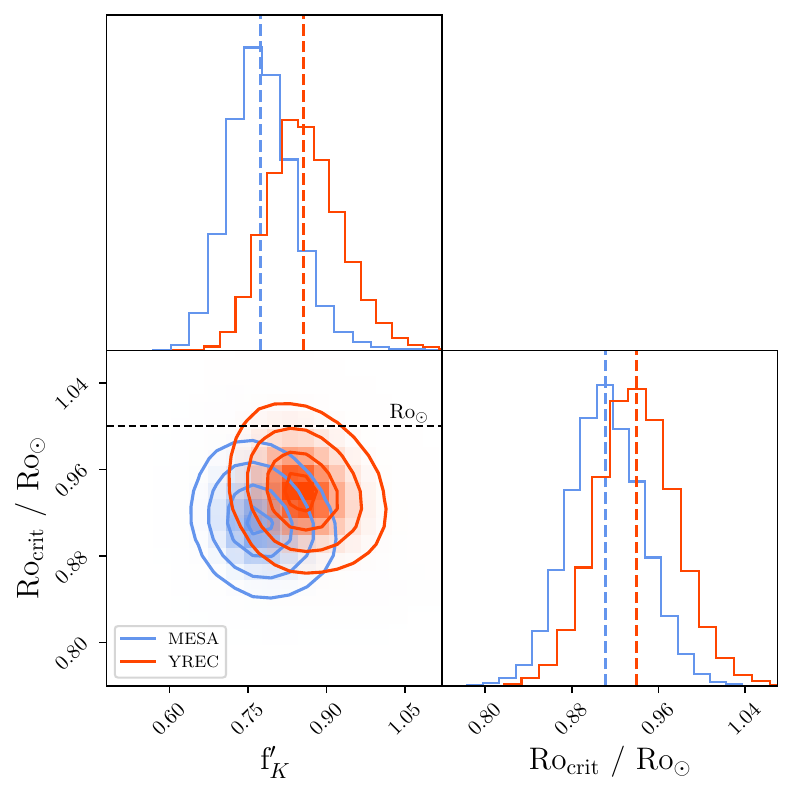}{.45\textwidth}{(a)}
              \fig{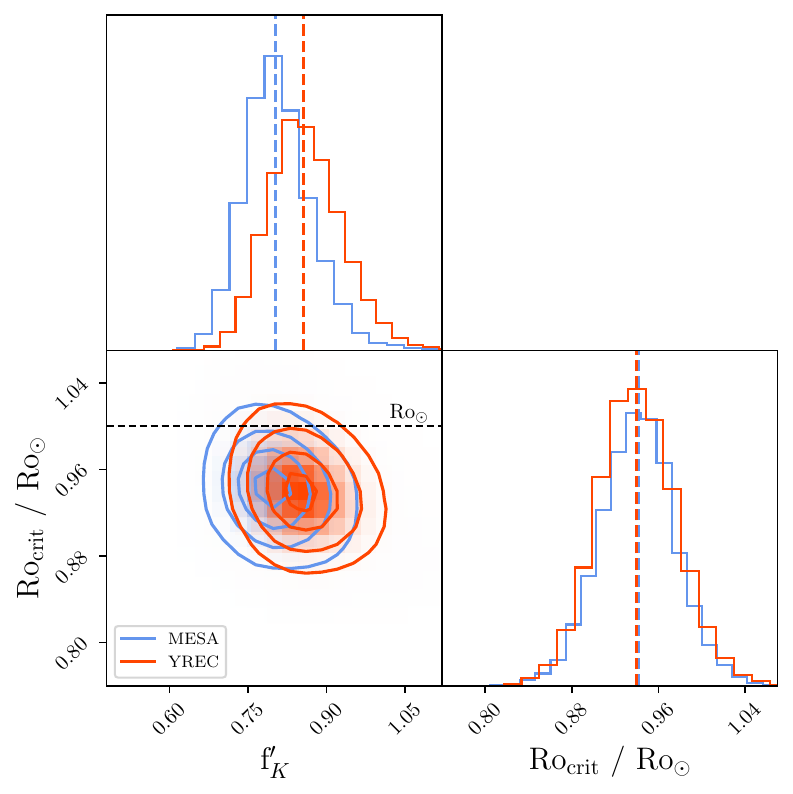}{.45\textwidth}{(b)}}
    \caption{\textbf{(a)} Corner plot showing the marginal and joint posterior distributions for the global parameters of our WMB model. Blue shows the samples from the fit using a neural net trained on a grid of \mesa models, and red shows the samples from a fit using the \yrec-trained neural net. The solar Rossby number Ro$_\odot$ is shown as a dashed black line. The median values of each distribution are shown as dashed lines in their respective colors in the top and right panels. \textbf{(b)} The same posterior distributions, now with the He-enrichment law in the \mesa probabilistic model fixed to the relation used when generating the \yrec grid. The primary difference between the grids used to train the emulator models is the varied Helium abundance \yini in the \mesa grid. When fixed to the \yrec enrichment law, the constraints on WMB global parameters are in closer agreement.}
    \label{fig:corner}
\end{figure*}

The left panel of Figure \ref{fig:corner} shows the marginal and joint posterior distributions for the braking law parameters when fit with the \mesa and \yrec ANNs. The black dashed line shows the solar Rossby number, Ro$_\odot$. Both \mesa and \yrec return values of \rocrit below Ro$_\odot$, indicating that the onset of WMB occurs before the age of the Sun for a solar analog. The inferred braking law parameters have slight offsets, but agree within 1$\sigma$. To assess the impact of leaving the \yini parameter free, we also performed probabilistic modeling with the \mesa ANN with \yini set to the He-enrichment law described above. We show the updated posterior distributions for this fit compared to the \yrec ANN in the right panel of Figure \ref{fig:corner}.

Using the \yrec emulator model, we retrieve constraints on the braking law parameters of $f_K' = 0.86 \pm 0.07$ and \rocrit $=0.94 \pm 0.04$. When \yini is left as a free parameter, the \mesa emulator model returns $f_K' = 0.77 \pm 0.07$ and \rocrit $=0.91 \pm 0.03$. When we fix the Helium enrichment law to that used in the \yrec grid, the \mesa emulator model reports  $f_K' = 0.80 \pm 0.07$ and \rocrit $=0.94 \pm 0.03$. We note that all models consistently return a value of \rocrit below the solar Rossby number.

When holding \yini fixed to the \yrec He-enrichment law, we find closer agreement between the braking law parameters inferred by our model fitting, with \rocrit in near-perfect agreement. This implies that \yini provides additional constraints on the braking law parameters, and its inclusion as a grid dimension can influence the result. \yini is a challenging property to measure for sun-like stars, and yet affects our inferred value of \rocrit at the $\sim$1$\sigma$-level. We conclude that uncertainty in the Helium enrichment law should be treated as a systematic uncertainty in the inference of \rocrit.

\subsection{Future Applications} \label{sec:future}

In this study, we focus only on $f_K$ and \rocrit due to the age distribution of our sample. In the future, the same approach described here could be applied to a sample of targets which span earlier phases of evolution (i.e. young open clusters), at which time braking law assumptions, such as the disk-locking timescale, disk lifetime, $\omega_{\rm sat}$, and internal angular momentum transport must be treated more carefully.

We limited the range of our input model grid to cover the parameters of our sample in order to reduce the computational time required for model generation and neural network training. The framework for the ANN emulator could easily be applied to a grid spanning a wider range of stellar properties, and would provide a useful tool for quickly evaluating stellar evolution tracks or simulating stellar populations. To reduce training time, the grid resolution could be selectively increased to reach a precision threshold. \cite{scutt2023} suggested that parameter spacing can be modified in different regions of the grid to improve ANN precision. 

Asteroseismic pulsation frequencies are often generated alongside stellar models using tools such as \textsf{GYRE} \citep{townsend2013}. These pulsation frequencies, particularly the large frequency spacing ($\Delta \nu$), can be included in the grid dimensions (e.g. \citealt{lyttle2021}) and applied as further likelihood constraints for models. Ideally, some combination of the above additions could be implemented to produce a broadly applicable stellar evolution emulator that does not require generating or interpolating large model grids.

\section{Conclusions} \label{sec:conclusions}

In summary, our primary conclusions are:

\begin{enumerate}
    \item We present evidence for weakened magnetic braking in old stars. Using a neural network as a stellar evolution emulator, we perform probabilistic modeling to produce posterior distributions for the parameters of the weakened braking model. We find that the weakened braking model provides the best fit to the observed distribution of rotation periods.
    \item We show that the most likely weakened braking scenario diverges from standard spindown at a slightly earlier evolutionary phase than the Sun (\rocrit /Ro$_\odot=0.91\pm0.03$). We caution that our WMB model is a simplified case in which angular momentum loss is fully switched off at a critical Rossby number, and likely does not fully capture the time evolution of the stellar dynamo. The relatively sparse calibrator sample near \rocrit means that it remains challenging to infer the precise onset of WMB relative to the Sun's evolution.
    \item Our method for emulating stellar evolution with a neural network enables rapid evaluation of stellar models, making it possible to fit braking law parameters while properly accounting for the uncertainties in the stellar parameters of our calibrator sample. By modifying the braking law used to generate our training set, we could test other effects at early times, such as the impact of internal angular momentum transport or disk-locking. 
    \item We report mild disagreement between the constraints on WMB parameters when using different underlying model grids. This indicates that the choice of grid physics and which parameters are varied in the model can impact the inferred model parameters. For our choices, the impact is at the 1$\sigma$ level.
    \item The WMB model appears compatible with the solar twins sample. The standard spindown model predicts slower rotation than observed in the solar twins stars during the second half of the main sequence, while their rotation periods can be described by the WMB model with modest deviations from a fully edge-on population.
    \item Our constraint on the \rocrit at which stars enter a phase of weakened braking suggests that gyrochronology faces challenges when estimating stellar ages for much of the main sequence lifetime. For sun-like stars, gyrochronological age estimates are likely unreliable beyond an age of $\sim$4 Gyr. For more massive stars ($\gtrsim 1.1$M$_\odot$), gyrochronology relations appear break down even earlier, at an age of $\sim$2.5 Gyr. Even after a star has entered the weakened braking phase, a reasonable range for its age can be estimated from its rotation period, and our constraint on \rocrit enables gyrochronological modeling that will provide a realistic uncertainty on the stellar age.
\end{enumerate}

The growing population of stars with precisely measured ages and rotation periods from asteroseismology is shedding essential light on the evolution of stellar rotation. Improved direct observations of magnetic field strength can add additional constraints on the braking law parameters. As more stars are added to this sample, the transition to WMB can be constrained to higher precision. 

\section*{Acknowledgements}

N.S. acknowledges support by the National Science Foundation Graduate Research Fellowship Program under Grant No. 1842402. J.v.S. and N.S. acknowledge support from the Research Corporation for Science Advancement through Scialog award \#39436, in partnership with the Heising-Simons Foundation. J.v.S. also acknowledges support from the National Science Foundation grant AST-2205888. This work has received funding from the European Research Council (ERC) under the European Union’s Horizon 2020 research and innovation programme (CartographY GA. 804752). T.S.M.\ acknowledges support from NASA grant 80NSSC22K0475. Computational time at the Texas Advanced Computing Center was provided through XSEDE allocation TG-AST090107. A.J.L.\ acknowledges the support of the Science and Technology Facilities Council. R.H.D.T.\ acknowledges support from NASA grant 80NSSC20K0515.

\bibliography{references}{}

\begin{thebibliography}{}
\expandafter\ifx\csname natexlab\endcsname\relax\def\natexlab#1{#1}\fi
\providecommand{\url}[1]{\href{#1}{#1}}
\providecommand{\dodoi}[1]{doi:~\href{http://doi.org/#1}{\nolinkurl{#1}}}
\providecommand{\doeprint}[1]{\href{http://ascl.net/#1}{\nolinkurl{http://ascl.net/#1}}}
\providecommand{\doarXiv}[1]{\href{https://arxiv.org/abs/#1}{\nolinkurl{https://arxiv.org/abs/#1}}}

\bibitem[{Abadi {et~al.}(2016)Abadi, Barham, Chen, Chen, Davis, Dean, Devin,
  Ghemawat, Irving, Isard, Kudlur, Levenberg, Monga, Moore, Murray, Steiner,
  Tucker, Vasudevan, Warden, Wicke, Yu, \& Zheng}]{abadi2016}
Abadi, M., Barham, P., Chen, J., {et~al.} 2016, {{TensorFlow}}: {{A}} System
  for Large-Scale Machine Learning,  {arXiv}.
\newblock \doeprint{1605.08695}

\bibitem[{Adelberger {et~al.}(2011)Adelberger, Balantekin, Bemmerer, Bertulani,
  Chen, Costantini, Couder, Cyburt, Davids, Freedman, Gai, Garcia, Gazit,
  Gialanella, Greife, Hass, Heeger, Haxton, Imbriani, Itahashi, Junghans,
  Kubodera, Langanke, Leitner, Leitner, Marcucci, Motobayashi, Mukhamedzhanov,
  Nollett, Nunes, Park, Parker, Prati, {Ramsey-Musolf}, Robertson, Schiavilla,
  Simpson, Snover, Spitaleri, Strieder, Suemmerer, Trautvetter, Tribble, Typel,
  Uberseder, Vetter, Wiescher, \& Winslow}]{adelberger2011}
Adelberger, E.~G., Balantekin, A.~B., Bemmerer, D., {et~al.} 2011, Reviews of
  Modern Physics, 83, 195, \dodoi{10.1103/RevModPhys.83.195}

\bibitem[{Aguirre {et~al.}(2015)Aguirre, Davies, Basu, {Christensen-Dalsgaard},
  Creevey, Metcalfe, Bedding, Casagrande, Handberg, Lund, Nissen, Chaplin,
  Huber, Serenelli, Stello, Van~Eylen, Campante, Elsworth, Gilliland, Hekker,
  Karoff, Kawaler, Kjeldsen, \& Lundkvist}]{aguirre2015}
Aguirre, V.~S., Davies, G.~R., Basu, S., {et~al.} 2015, Monthly Notices of the
  Royal Astronomical Society, 452, 2127, \dodoi{10.1093/mnras/stv1388}

\bibitem[{Angus {et~al.}(2015)Angus, Aigrain, Foreman-Mackey, \&
  McQuillan}]{angus2015}
Angus, R., Aigrain, S., Foreman-Mackey, D., \& McQuillan, A. 2015, Monthly
  Notices of the Royal Astronomical Society, 450, 1787,
  \dodoi{10.1093/mnras/stv423}

\bibitem[{Angus {et~al.}(2019)Angus, Morton, Foreman-Mackey, van Saders,
  Curtis, Kane, Bedell, Kiman, Hogg, \& Brewer}]{angus2019}
Angus, R., Morton, T.~D., Foreman-Mackey, D., {et~al.} 2019, The Astronomical
  Journal, 158, 173, \dodoi{10.3847/1538-3881/ab3c53}

\bibitem[{{Ball} \& {Gizon}(2014)}]{Ball2014}
{Ball}, W.~H., \& {Gizon}, L. 2014, \aap, 568, A123,
  \dodoi{10.1051/0004-6361/201424325}

\bibitem[{Barnes(2007)}]{barnes2007}
Barnes, S.~A. 2007, The Astrophysical Journal, 669, 1167,
  \dodoi{10.1086/519295}

\bibitem[{Barnes(2010)}]{barnes2010}
---. 2010, The Astrophysical Journal, 722, 222,
  \dodoi{10.1088/0004-637X/722/1/222}

\bibitem[{Berger {et~al.}(2020)Berger, Huber, van Saders, Gaidos, Tayar, \&
  Kraus}]{berger2020}
Berger, T.~A., Huber, D., van Saders, J.~L., {et~al.} 2020, The Astronomical
  Journal, 159, 280, \dodoi{10.3847/1538-3881/159/6/280}

\bibitem[{Borucki {et~al.}(2010)Borucki, Koch, Basri, Batalha, Brown, Caldwell,
  Caldwell, {Christensen-Dalsgaard}, Cochran, DeVore, Dunham, Dupree, Gautier,
  Geary, Gilliland, Gould, Howell, Jenkins, Kondo, Latham, Marcy, Meibom,
  Kjeldsen, Lissauer, Monet, Morrison, Sasselov, Tarter, Boss, Brownlee, Owen,
  Buzasi, Charbonneau, Doyle, Fortney, Ford, Holman, Seager, Steffen, Welsh,
  Rowe, Anderson, Buchhave, Ciardi, Walkowicz, Sherry, Horch, Isaacson,
  Everett, Fischer, Torres, Johnson, Endl, MacQueen, Bryson, Dotson, Haas,
  Kolodziejczak, Van~Cleve, Chandrasekaran, Twicken, Quintana, Clarke, Allen,
  Li, Wu, Tenenbaum, Verner, Bruhweiler, Barnes, \& Prsa}]{borucki2010}
Borucki, W.~J., Koch, D., Basri, G., {et~al.} 2010, Science, 327, 977,
  \dodoi{10.1126/science.1185402}

\bibitem[{Chaplin {et~al.}(2011)Chaplin, Bedding, Bonanno, Broomhall, Garcia,
  Hekker, Huber, Verner, Basu, Elsworth, Houdek, Mathur, Mosser, New, Stevens,
  Appourchaux, Karoff, Metcalfe, {Molenda-Zakowicz}, Monteiro, Thompson,
  {Christensen-Dalsgaard}, Gilliland, Kawaler, Kjeldsen, Ballot, Benomar,
  Corsaro, Campante, Gaulme, Hale, Handberg, Jarvis, Regulo, Roxburgh,
  Salabert, Stello, Mullally, Li, \& Wohler}]{chaplin2011}
Chaplin, W.~J., Bedding, T.~R., Bonanno, A., {et~al.} 2011, The Astrophysical
  Journal, 732, L5, \dodoi{10.1088/2041-8205/732/1/L5}

\bibitem[{Cody {et~al.}(2018)Cody, Barentsen, Hedges, {Gully-Santiago}, Dotson,
  Barclay, Bryson, \& Saunders}]{codyCatalog29Open2018}
Cody, A.~M., Barentsen, G., Hedges, C., {et~al.} 2018, Research Notes of the
  AAS, 2, 199, \dodoi{10.3847/2515-5172/aaec76}

\bibitem[{{Cox} \& {Giuli}(1968)}]{cox1968}
{Cox}, J.~P., \& {Giuli}, R.~T. 1968, {Principles of stellar structure}

\bibitem[{Creevey {et~al.}(2017)Creevey, Metcalfe, Schultheis, Salabert, Bazot,
  Thevenin, Mathur, Xu, \& Garcia}]{creevey2017}
Creevey, O., Metcalfe, T.~S., Schultheis, M., {et~al.} 2017, Astronomy \&
  Astrophysics, 601, A67, \dodoi{10.1051/0004-6361/201629496}

\bibitem[{Curtis {et~al.}(2019)Curtis, Ag{\"u}eros, Douglas, \&
  Meibom}]{curtis2019}
Curtis, J.~L., Ag{\"u}eros, M.~A., Douglas, S.~T., \& Meibom, S. 2019, The
  Astrophysical Journal, 879, 49, \dodoi{10.3847/1538-4357/ab2393}

\bibitem[{Curtis {et~al.}(2020)Curtis, Ag{\"u}eros, Matt, Covey, Douglas,
  Angus, Saar, Cody, Vanderburg, Law, Kraus, Latham, Baranec, Riddle, Ziegler,
  Lund, Torres, Meibom, Aguirre, \& Wright}]{curtis2020}
Curtis, J.~L., Ag{\"u}eros, M.~A., Matt, S.~P., {et~al.} 2020, The
  Astrophysical Journal, 904, 140, \dodoi{10.3847/1538-4357/abbf58}

\bibitem[{David {et~al.}(2022)David, Angus, Curtis, {van Saders}, Colman,
  Contardo, Lu, \& Zinn}]{david2022a}
David, T.~J., Angus, R., Curtis, J.~L., {et~al.} 2022, The Astrophysical
  Journal, 933, 114, \dodoi{10.3847/1538-4357/ac6dd3}

\bibitem[{Davies {et~al.}(2015)Davies, Chaplin, Farr, García, Lund, Mathis,
  Metcalfe, Appourchaux, Basu, Benomar, Campante, Ceillier, Elsworth, Handberg,
  Salabert, \& Stello}]{davies2015}
Davies, G.~R., Chaplin, W.~J., Farr, W.~M., {et~al.} 2015, Monthly Notices of
  the Royal Astronomical Society, 446, 2959, \dodoi{10.1093/mnras/stu2331}

\bibitem[{Davies {et~al.}(2016)Davies, Aguirre, Bedding, Handberg, Lund,
  Chaplin, Huber, White, Benomar, Hekker, Basu, Campante,
  {Christensen-Dalsgaard}, Elsworth, Karoff, Kjeldsen, Lundkvist, Metcalfe, \&
  Stello}]{davies2016}
Davies, G.~R., Aguirre, V.~S., Bedding, T.~R., {et~al.} 2016, Monthly Notices
  of the Royal Astronomical Society, 456, 2183, \dodoi{10.1093/mnras/stv2593}

\bibitem[{Demarque {et~al.}(2008)Demarque, Guenther, Li, Mazumdar, \&
  Straka}]{demarque2008a}
Demarque, P., Guenther, D.~B., Li, L.~H., Mazumdar, A., \& Straka, C.~W. 2008,
  Astrophysics and Space Science, 316, 31, \dodoi{10.1007/s10509-007-9698-y}

\bibitem[{Denissenkov {et~al.}(2010)Denissenkov, Pinsonneault, Terndrup, \&
  Newsham}]{denissenkov2010}
Denissenkov, P.~A., Pinsonneault, M., Terndrup, D.~M., \& Newsham, G. 2010, The
  Astrophysical Journal, 716, 1269, \dodoi{10.1088/0004-637X/716/2/1269}

\bibitem[{dos Santos {et~al.}(2016)dos Santos, Mel{\'e}ndez, do~Nascimento,
  Bedell, Ram{\'i}rez, Bean, Asplund, Spina, Dreizler, {Alves-Brito}, \&
  Casagrande}]{santos2016}
dos Santos, L.~A., Mel{\'e}ndez, J., do~Nascimento, J.-D., {et~al.} 2016,
  Astronomy \& Astrophysics, 592, A156, \dodoi{10.1051/0004-6361/201628558}

\bibitem[{Dungee {et~al.}(2022)Dungee, {van Saders}, Gaidos, Chun, Garcia,
  Magnier, Mathur, \& Santos}]{dungee2022}
Dungee, R., {van Saders}, J., Gaidos, E., {et~al.} 2022, The Astrophysical
  Journal, 938, 118, \dodoi{10.3847/1538-4357/ac90be}

\bibitem[{Epstein \& Pinsonneault(2013)}]{epstein2013}
Epstein, C.~R., \& Pinsonneault, M.~H. 2013, The Astrophysical Journal, 780,
  159, \dodoi{10.1088/0004-637X/780/2/159}

\bibitem[{{Gaia Collaboration} {et~al.}(2018){Gaia Collaboration}, Brown,
  Vallenari, Prusti, {de Bruijne}, Babusiaux, {Bailer-Jones}, Biermann, Evans,
  Eyer, Jansen, Jordi, Klioner, Lammers, Lindegren, Luri, Mignard, Panem,
  Pourbaix, Randich, Sartoretti, Siddiqui, Soubiran, {van Leeuwen}, Walton,
  Arenou, Bastian, Cropper, Drimmel, Katz, Lattanzi, Bakker, Cacciari,
  Casta{\~n}eda, Chaoul, Cheek, De~Angeli, Fabricius, Guerra, Holl, Masana,
  Messineo, Mowlavi, Nienartowicz, Panuzzo, Portell, Riello, Seabroke, Tanga,
  Th{\'e}venin, {Gracia-Abril}, Comoretto, {Garcia-Reinaldos}, Teyssier,
  Altmann, Andrae, Audard, {Bellas-Velidis}, Benson, Berthier, Blomme, Burgess,
  Busso, Carry, Cellino, Clementini, Clotet, Creevey, Davidson, De~Ridder,
  Delchambre, Dell'Oro, Ducourant, {Fern{\'a}ndez-Hern{\'a}ndez}, Fouesneau,
  Fr{\'e}mat, Galluccio, {Garc{\'i}a-Torres}, {Gonz{\'a}lez-N{\'u}{\~n}ez},
  {Gonz{\'a}lez-Vidal}, Gosset, Guy, Halbwachs, Hambly, Harrison,
  Hern{\'a}ndez, Hestroffer, Hodgkin, Hutton, Jasniewicz,
  {Jean-Antoine-Piccolo}, Jordan, Korn, {Krone-Martins}, Lanzafame, Lebzelter,
  L{\"o}ffler, Manteiga, Marrese, {Mart{\'i}n-Fleitas}, Moitinho, Mora,
  Muinonen, Osinde, Pancino, Pauwels, Petit, {Recio-Blanco}, Richards,
  Rimoldini, Robin, Sarro, Siopis, Smith, Sozzetti, S{\"u}veges, Torra, {van
  Reeven}, Abbas, Abreu~Aramburu, Accart, Aerts, Altavilla, {\'A}lvarez,
  Alvarez, Alves, Anderson, Andrei, Anglada~Varela, Antiche, Antoja, Arcay,
  Astraatmadja, Bach, Baker, {Balaguer-N{\'u}{\~n}ez}, Balm, Barache, Barata,
  Barbato, Barblan, Barklem, Barrado, Barros, Barstow,
  Bartholom{\'e}~Mu{\~n}oz, Bassilana, Becciani, Bellazzini, Berihuete,
  Bertone, Bianchi, Bienaym{\'e}, {Blanco-Cuaresma}, Boch, Boeche, Bombrun,
  Borrachero, Bossini, Bouquillon, Bourda, Bragaglia, Bramante, Breddels,
  Bressan, Brouillet, Br{\"u}semeister, Brugaletta, Bucciarelli, Burlacu,
  Busonero, Butkevich, Buzzi, Caffau, Cancelliere, Cannizzaro, {Cantat-Gaudin},
  Carballo, Carlucci, Carrasco, Casamiquela, Castellani, {Castro-Ginard},
  Charlot, Chemin, Chiavassa, Cocozza, Costigan, Cowell, Crifo, Crosta,
  Crowley, Cuypers{\textdagger}, Dafonte, Damerdji, Dapergolas, David, David,
  {de Laverny}, De~Luise, De~March, {de Martino}, {de Souza}, {de Torres},
  Debosscher, {del Pozo}, Delbo, Delgado, Delgado, Di~Matteo, Diakite, Diener,
  Distefano, Dolding, Drazinos, Dur{\'a}n, Edvardsson, Enke, Eriksson, Esquej,
  Eynard~Bontemps, Fabre, Fabrizio, Faigler, Falc{\~a}o, Farr{\`a}s~Casas,
  Federici, Fedorets, Fernique, Figueras, Filippi, Findeisen, Fonti, Fraile,
  Fraser, Fr{\'e}zouls, Gai, Galleti, Garabato, {Garc{\'i}a-Sedano}, Garofalo,
  Garralda, Gavel, Gavras, Gerssen, Geyer, Giacobbe, Gilmore, Girona,
  Giuffrida, Glass, Gomes, Granvik, Gueguen, Guerrier, Guiraud,
  {Guti{\'e}rrez-S{\'a}nchez}, Haigron, Hatzidimitriou, Hauser, Haywood,
  Heiter, Helmi, Heu, Hilger, Hobbs, Hofmann, Holland, Huckle, Hypki, Icardi,
  Jan{\ss}en, {Jevardat de Fombelle}, Jonker, Juh{\'a}sz, Julbe, Karampelas,
  Kewley, Klar, Kochoska, Kohley, Kolenberg, Kontizas, Kontizas, Koposov,
  Kordopatis, {Kostrzewa-Rutkowska}, Koubsky, Lambert, Lanza, Lasne, Lavigne,
  Le~Fustec, {Le Poncin-Lafitte}, Lebreton, Leccia, Leclerc, {Lecoeur-Taibi},
  Lenhardt, Leroux, Liao, Licata, Lindstr{\o}m, Lister, Livanou, Lobel,
  L{\'o}pez, Managau, Mann, Mantelet, Marchal, Marchant, Marconi, Marinoni,
  Marschalk{\'o}, Marshall, Martino, Marton, Mary, Massari, Matijevi{\v c},
  Mazeh, McMillan, Messina, Michalik, Millar, Molina, Molinaro, Moln{\'a}r,
  Montegriffo, Mor, Morbidelli, Morel, Morris, Mulone, Muraveva, Musella,
  Nelemans, Nicastro, Noval, O'Mullane, Ord{\'e}novic,
  {Ord{\'o}{\~n}ez-Blanco}, Osborne, Pagani, Pagano, Pailler, Palacin,
  Palaversa, Panahi, Pawlak, Piersimoni, Pineau, Plachy, Plum, Poggio,
  Poujoulet, Pr{\v s}a, Pulone, Racero, Ragaini, Rambaux, {Ramos-Lerate},
  Regibo, Reyl{\'e}, Riclet, Ripepi, Riva, Rivard, Rixon, Roegiers, Roelens,
  {Romero-G{\'o}mez}, Rowell, Royer, {Ruiz-Dern}, Sadowski,
  Sagrist{\`a}~Sell{\'e}s, Sahlmann, Salgado, Salguero, Sanna, {Santana-Ros},
  Sarasso, Savietto, Schultheis, Sciacca, Segol, Segovia, S{\'e}gransan, Shih,
  Siltala, Silva, Smart, Smith, Solano, Solitro, Sordo, Soria~Nieto, Souchay,
  Spagna, Spoto, Stampa, Steele, Steidelm{\"u}ller, Stephenson, Stoev, Suess,
  Surdej, Szabados, {Szegedi-Elek}, Tapiador, Taris, Tauran, Taylor, Teixeira,
  Terrett, Teyssandier, Thuillot, Titarenko, Torra~Clotet, Turon, Ulla,
  Utrilla, Uzzi, Vaillant, Valentini, Valette, {van Elteren}, Van~Hemelryck,
  {van Leeuwen}, Vaschetto, Vecchiato, Veljanoski, Viala, Vicente, Vogt, {von
  Essen}, Voss, Votruba, Voutsinas, Walmsley, Weiler, Wertz, Wevers,
  Wyrzykowski, Yoldas, {\v Z}erjal, Ziaeepour, Zorec, Zschocke, Zucker,
  Zurbach, \& Zwitter}]{gaiacollaboration2018}
{Gaia Collaboration}, Brown, A. G.~A., Vallenari, A., {et~al.} 2018, Astronomy
  \& Astrophysics, 616, A1, \dodoi{10.1051/0004-6361/201833051}

\bibitem[{Gallet \& Bouvier(2013)}]{gallet2013}
Gallet, F., \& Bouvier, J. 2013, Astronomy \& Astrophysics, 556, A36,
  \dodoi{10.1051/0004-6361/201321302}

\bibitem[{Gallet \& Bouvier(2015)}]{gallet2015}
---. 2015, Astronomy \& Astrophysics, 577, A98,
  \dodoi{10.1051/0004-6361/201525660}

\bibitem[{Garraffo {et~al.}(2016)Garraffo, Drake, \& Cohen}]{garraffo2016}
Garraffo, C., Drake, J.~J., \& Cohen, O. 2016, Astronomy \& Astrophysics, 595,
  A110, \dodoi{10.1051/0004-6361/201628367}

\bibitem[{Gelman \& Rubin(1992)}]{gelman1992}
Gelman, A., \& Rubin, D.~B. 1992, Statistical Science, 7, 457,
  \dodoi{10.1214/ss/1177011136}

\bibitem[{Grevesse \& Sauval(1998)}]{grevesse1998}
Grevesse, N., \& Sauval, A. 1998, Space Science Reviews, 85, 161,
  \dodoi{10.1023/A:1005161325181}

\bibitem[{Hall {et~al.}(2021)Hall, Davies, {van Saders}, Nielsen, Lund,
  Chaplin, Garc{\'i}a, Amard, Breimann, Khan, See, \& Tayar}]{hall2021}
Hall, O.~J., Davies, G.~R., {van Saders}, J., {et~al.} 2021, Nature Astronomy,
  \dodoi{10.1038/s41550-021-01335-x}

\bibitem[{Howell {et~al.}(2014)Howell, Sobeck, Haas, Still, Barclay, Mullally,
  Troeltzsch, Aigrain, Bryson, Caldwell, Chaplin, Cochran, Huber, Marcy,
  Miglio, Najita, Smith, Twicken, \&
  Fortney}]{howellK2MissionCharacterization2014}
Howell, S., Sobeck, C., Haas, M.~R., {et~al.} 2014, Publications of the
  Astronomical Society of the Pacific, 126, 398, \dodoi{10.1086/676406}

\bibitem[{Hoﬀman \& Gelman(2014)}]{nuts}
Hoﬀman, M.~D., \& Gelman, A. 2014, Journal of Machine Learning Research, 15,
  1351

\bibitem[{Huber {et~al.}(2011)Huber, Bedding, Stello, Hekker, Mathur, Mosser,
  Verner, Bonanno, Buzasi, Campante, Elsworth, Hale, Kallinger, Aguirre,
  Chaplin, De~Ridder, Garcia, Appourchaux, Frandsen, Houdek,
  {Molenda-Zakowicz}, Monteiro, {Christensen-Dalsgaard}, Gilliland, Kawaler,
  Kjeldsen, Broomhall, Corsaro, Salabert, Sanderfer, Seader, \&
  Smith}]{huber2011}
Huber, D., Bedding, T.~R., Stello, D., {et~al.} 2011, The Astrophysical
  Journal, 743, 143, \dodoi{10.1088/0004-637X/743/2/143}

\bibitem[{Kawaler(1988)}]{kawaler1988}
Kawaler, S.~D. 1988, The Astrophysical Journal, 333, 236,
  \dodoi{10.1086/166740}

\bibitem[{{Kim} \& {Demarque}(1996)}]{kim1996}
{Kim}, Y.-C., \& {Demarque}, P. 1996, \apj, 457, 340, \dodoi{10.1086/176733}

\bibitem[{Kingma \& Ba(2017)}]{kingma2017}
Kingma, D.~P., \& Ba, J. 2017, Adam: {{A Method}} for {{Stochastic
  Optimization}},  {arXiv}.
\newblock \doeprint{1412.6980}

\bibitem[{Krishnamurthi {et~al.}(1997)Krishnamurthi, Pinsonneault, Barnes, \&
  Sofia}]{krishnamurthi1997}
Krishnamurthi, A., Pinsonneault, M.~H., Barnes, S., \& Sofia, S. 1997, The
  Astrophysical Journal, 480, 303, \dodoi{10.1086/303958}

\bibitem[{{Kurucz}(1997)}]{kurucz1997}
{Kurucz}, R.~L. 1997, in The Third Conference on Faint Blue Stars, ed. A.~G.~D.
  {Philip}, J.~{Liebert}, R.~{Saffer}, \& D.~S. {Hayes}, 33

\bibitem[{{Lorenzo-Oliveira} {et~al.}(2019){Lorenzo-Oliveira}, Mel{\'e}ndez,
  Galarza, Ponte, dos Santos, Spina, Bedell, Ram{\'i}rez, Bean, \&
  Asplund}]{lorenzo-oliveira2019}
{Lorenzo-Oliveira}, D., Mel{\'e}ndez, J., Galarza, J.~Y., {et~al.} 2019,
  Monthly Notices of the Royal Astronomical Society: Letters, 485, L68,
  \dodoi{10.1093/mnrasl/slz034}

\bibitem[{Lund {et~al.}(2017)Lund, Aguirre, Davies, Chaplin,
  {Christensen-Dalsgaard}, Houdek, White, Bedding, Ball, Huber, Antia,
  Lebreton, Latham, Handberg, Verma, Basu, Casagrande, Justesen, Kjeldsen, \&
  Mosumgaard}]{lund2017}
Lund, M.~N., Aguirre, V.~S., Davies, G.~R., {et~al.} 2017, The Astrophysical
  Journal, 835, 172, \dodoi{10.3847/1538-4357/835/2/172}

\bibitem[{Lyttle {et~al.}(2021)Lyttle, Davies, Li, Carboneau, Leung, Westwood,
  Chaplin, Hall, Huber, Nielsen, Basu, \& Garc{\'i}a}]{lyttle2021}
Lyttle, A.~J., Davies, G.~R., Li, T., {et~al.} 2021, arXiv:2105.04482
  [astro-ph].
\newblock \doarXiv{2105.04482}

\bibitem[{Mamajek \& Hillenbrand(2008)}]{mamajek2008a}
Mamajek, E.~E., \& Hillenbrand, L.~A. 2008, The Astrophysical Journal, 687,
  1264, \dodoi{10.1086/591785}

\bibitem[{Matt {et~al.}(2015)Matt, Brun, Baraffe, Bouvier, \&
  Chabrier}]{matt2015}
Matt, S.~P., Brun, A.~S., Baraffe, I., Bouvier, J., \& Chabrier, G. 2015, The
  Astrophysical Journal Letters, 799, L23, \dodoi{10.1088/2041-8205/799/2/L23}

\bibitem[{Matt {et~al.}(2012)Matt, MacGregor, Pinsonneault, \&
  Greene}]{matt2012}
Matt, S.~P., MacGregor, K.~B., Pinsonneault, M.~H., \& Greene, T.~P. 2012, The
  Astrophysical Journal, 754, L26, \dodoi{10.1088/2041-8205/754/2/L26}

\bibitem[{McQuillan {et~al.}(2014)McQuillan, Mazeh, \&
  Aigrain}]{mcquillan2014a}
McQuillan, A., Mazeh, T., \& Aigrain, S. 2014, The Astrophysical Journal
  Supplement Series, 211, 24, \dodoi{10.1088/0067-0049/211/2/24}

\bibitem[{Meibom {et~al.}(2015)Meibom, Barnes, Platais, Gilliland, Latham, \&
  Mathieu}]{meibom2015}
Meibom, S., Barnes, S.~A., Platais, I., {et~al.} 2015, Nature, 517, 589,
  \dodoi{10.1038/nature14118}

\bibitem[{Meibom {et~al.}(2011)Meibom, Barnes, Latham, Batalha, Borucki, Koch,
  Basri, Walkowicz, Janes, Jenkins, Van~Cleve, Haas, Bryson, Dupree, Furesz,
  Szentgyorgyi, Buchhave, Clarke, Twicken, \& Quintana}]{meibom2011}
Meibom, S., Barnes, S.~A., Latham, D.~W., {et~al.} 2011, The Astrophysical
  Journal, 733, L9, \dodoi{10.1088/2041-8205/733/1/L9}

\bibitem[{Mendoza {et~al.}(2007)Mendoza, Seaton, Buerger, Bellor{\'i}n,
  Mel{\'e}ndez, Gonz{\'a}lez, Rodr{\'i}guez, Delahaye, Palacios, Pradhan, \&
  Zeippen}]{mendoza2007}
Mendoza, C., Seaton, M.~J., Buerger, P., {et~al.} 2007, Monthly Notices of the
  Royal Astronomical Society, 378, 1031,
  \dodoi{10.1111/j.1365-2966.2007.11837.x}

\bibitem[{{Metcalfe} \& {Charbonneau}(2003)}]{Metcalfe2003}
{Metcalfe}, T.~S., \& {Charbonneau}, P. 2003, Journal of Computational Physics,
  185, 176, \dodoi{10.1016/S0021-9991(02)00053-0}

\bibitem[{{Metcalfe} {et~al.}(2009){Metcalfe}, {Creevey}, \&
  {Christensen-Dalsgaard}}]{Metcalfe2009}
{Metcalfe}, T.~S., {Creevey}, O.~L., \& {Christensen-Dalsgaard}, J. 2009, \apj,
  699, 373, \dodoi{10.1088/0004-637X/699/1/373}

\bibitem[{Metcalfe {et~al.}(2016)Metcalfe, Egeland, \& van
  Saders}]{metcalfe2016}
Metcalfe, T.~S., Egeland, R., \& van Saders, J. 2016, The Astrophysical
  Journal, 826, L2, \dodoi{10.3847/2041-8205/826/1/L2}

\bibitem[{Metcalfe {et~al.}(2019)Metcalfe, Kochukhov, Ilyin, Strassmeier,
  {Godoy-Rivera}, \& Pinsonneault}]{metcalfe2019}
Metcalfe, T.~S., Kochukhov, O., Ilyin, I.~V., {et~al.} 2019, The Astrophysical
  Journal Letters, 887, L38, \dodoi{10.3847/2041-8213/ab5e48}

\bibitem[{{Metcalfe} {et~al.}(2023){Metcalfe}, {Townsend}, \&
  {Ball}}]{Metcalfe2023}
{Metcalfe}, T.~S., {Townsend}, R. H.~D., \& {Ball}, W.~H. 2023, RNAAS,
  accepted, \dodoi{10.48550/arXiv.2307.16247}

\bibitem[{Metcalfe {et~al.}(2014)Metcalfe, Creevey, Dogan, Mathur, Xu, Bedding,
  Chaplin, {Christensen-Dalsgaard}, Karoff, Trampedach, Benomar, Brown, Buzasi,
  Campante, Celik, Cunha, Davies, Deheuvels, Derekas, Di~Mauro, Garcia, Guzik,
  Howe, MacGregor, Mazumdar, Montalban, Monteiro, Salabert, Serenelli, Stello,
  Steslicki, Suran, Yildiz, Aksoy, Elsworth, Gruberbauer, Guenther, Lebreton,
  Molaverdikhani, Pricopi, Simoniello, \& White}]{metcalfe2014a}
Metcalfe, T.~S., Creevey, O.~L., Dogan, G., {et~al.} 2014, The Astrophysical
  Journal Supplement Series, 214, 27, \dodoi{10.1088/0067-0049/214/2/27}

\bibitem[{Metcalfe {et~al.}(2020)Metcalfe, {van Saders}, Basu, Buzasi, Chaplin,
  Egeland, Garcia, Gaulme, Huber, Reinhold, Schunker, Stassun, Appourchaux,
  Ball, Bedding, Deheuvels, {Gonzalez-Cuesta}, Handberg, Jimenez, Kjeldsen, Li,
  Lund, Mathur, Mosser, Nielsen, Noll, Orhan, Ortel, Santos, Yildiz, Baliunas,
  \& Soon}]{metcalfe2020}
Metcalfe, T.~S., {van Saders}, J.~L., Basu, S., {et~al.} 2020, The
  Astrophysical Journal, 900, 154, \dodoi{10.3847/1538-4357/aba963}

\bibitem[{Nielsen {et~al.}(2015)Nielsen, Schunker, Gizon, \&
  Ball}]{nielsen2015}
Nielsen, M.~B., Schunker, H., Gizon, L., \& Ball, W.~H. 2015, Astronomy \&
  Astrophysics, 582, A10, \dodoi{10.1051/0004-6361/201526615}

\bibitem[{Parker(1958)}]{parker1958}
Parker, E.~N. 1958, The Astrophysical Journal, 128, 664, \dodoi{10.1086/146579}

\bibitem[{Paxton {et~al.}(2010)Paxton, Bildsten, Dotter, Herwig, Lesaffre, \&
  Timmes}]{paxton2010}
Paxton, B., Bildsten, L., Dotter, A., {et~al.} 2010, The Astrophysical Journal
  Supplement Series, 192, 3, \dodoi{10.1088/0067-0049/192/1/3}

\bibitem[{Paxton {et~al.}(2013)Paxton, Cantiello, Arras, Bildsten, Brown,
  Dotter, Mankovich, Montgomery, Stello, Timmes, \& Townsend}]{paxton2013}
Paxton, B., Cantiello, M., Arras, P., {et~al.} 2013, The Astrophysical Journal
  Supplement Series, 208, 4, \dodoi{10.1088/0067-0049/208/1/4}

\bibitem[{Paxton {et~al.}(2015)Paxton, Marchant, Schwab, Bauer, Bildsten,
  Cantiello, Dessart, Farmer, Hu, Langer, Townsend, Townsley, \&
  Timmes}]{paxton2015}
Paxton, B., Marchant, P., Schwab, J., {et~al.} 2015, The Astrophysical Journal
  Supplement Series, 220, 15, \dodoi{10.1088/0067-0049/220/1/15}

\bibitem[{Paxton {et~al.}(2018)Paxton, Schwab, Bauer, Bildsten, Blinnikov,
  Duffell, Farmer, Goldberg, Marchant, Sorokina, Thoul, Townsend, \&
  Timmes}]{paxton2018}
Paxton, B., Schwab, J., Bauer, E.~B., {et~al.} 2018, The Astrophysical Journal
  Supplement Series, 234, 34, \dodoi{10.3847/1538-4365/aaa5a8}

\bibitem[{Paxton {et~al.}(2019)Paxton, Smolec, Schwab, Gautschy, Bildsten,
  Cantiello, Dotter, Farmer, Goldberg, Jermyn, Kanbur, Marchant, Thoul,
  Townsend, Wolf, Zhang, \& Timmes}]{paxton2019}
Paxton, B., Smolec, R., Schwab, J., {et~al.} 2019, The Astrophysical Journal
  Supplement Series, 243, 10, \dodoi{10.3847/1538-4365/ab2241}

\bibitem[{{Pinsonneault} {et~al.}(1989){Pinsonneault}, {Kawaler}, {Sofia}, \&
  {Demarque}}]{pinsonneault1989}
{Pinsonneault}, M.~H., {Kawaler}, S.~D., {Sofia}, S., \& {Demarque}, P. 1989,
  \apj, 338, 424, \dodoi{10.1086/167210}

\bibitem[{Rebull {et~al.}(2017)Rebull, Stauffer, Hillenbrand, Cody, Bouvier,
  Soderblom, Pinsonneault, \& Hebb}]{rebull2017}
Rebull, L.~M., Stauffer, J.~R., Hillenbrand, L.~A., {et~al.} 2017, The
  Astrophysical Journal, 839, 92, \dodoi{10.3847/1538-4357/aa6aa4}

\bibitem[{Reiners \& Mohanty(2012)}]{reiners2012}
Reiners, A., \& Mohanty, S. 2012, The Astrophysical Journal, 746, 43,
  \dodoi{10.1088/0004-637X/746/1/43}

\bibitem[{Reinhold {et~al.}(2020)Reinhold, Shapiro, Solanki, Montet, Krivova,
  Cameron, \& {Amazo-G{\'o}mez}}]{reinhold2020}
Reinhold, T., Shapiro, A.~I., Solanki, S.~K., {et~al.} 2020, Science, 368, 518,
  \dodoi{10.1126/science.aay3821}

\bibitem[{R{\'e}ville {et~al.}(2015)R{\'e}ville, Brun, Matt, Strugarek, \&
  Pinto}]{reville2015}
R{\'e}ville, V., Brun, A.~S., Matt, S.~P., Strugarek, A., \& Pinto, R.~F. 2015,
  The Astrophysical Journal, 798, 116, \dodoi{10.1088/0004-637X/798/2/116}

\bibitem[{{Rogers} \& {Nayfonov}(2002)}]{rogers2002}
{Rogers}, F.~J., \& {Nayfonov}, A. 2002, \apj, 576, 1064,
  \dodoi{10.1086/341894}

\bibitem[{{Rogers} {et~al.}(1996){Rogers}, {Swenson}, \&
  {Iglesias}}]{rogers1996}
{Rogers}, F.~J., {Swenson}, F.~J., \& {Iglesias}, C.~A. 1996, \apj, 456, 902,
  \dodoi{10.1086/176705}

\bibitem[{Salvatier {et~al.}(2016)Salvatier, Wiecki, \&
  Fonnesbeck}]{salvatier2016}
Salvatier, J., Wiecki, T.~V., \& Fonnesbeck, C. 2016, PeerJ Computer Science,
  2, e55, \dodoi{10.7717/peerj-cs.55}

\bibitem[{Santos {et~al.}(2021)Santos, Breton, Mathur, \&
  Garc{\'i}a}]{santos2021a}
Santos, A. R.~G., Breton, S.~N., Mathur, S., \& Garc{\'i}a, R.~A. 2021, The
  Astrophysical Journal Supplement Series, 255, 17,
  \dodoi{10.3847/1538-4365/ac033f}

\bibitem[{Scutt {et~al.}(2023)Scutt, Murphy, Nielsen, Davies, Bedding, \&
  Lyttle}]{scutt2023}
Scutt, O.~J., Murphy, S.~J., Nielsen, M.~B., {et~al.} 2023, Asteroseismology of
  \$\textbackslash delta\$ {{Scuti}} Stars: Emulating Model Grids Using a
  Neural Network,  {arXiv}.
\newblock \doeprint{2302.11025}

\bibitem[{Sills {et~al.}(2000)Sills, Pinsonneault, \& Terndrup}]{sills2000}
Sills, A., Pinsonneault, M.~H., \& Terndrup, D.~M. 2000, The Astrophysical
  Journal, 534, 335, \dodoi{10.1086/308739}

\bibitem[{Silva~Aguirre {et~al.}(2015)Silva~Aguirre, Davies, Basu,
  {Christensen-Dalsgaard}, Creevey, Metcalfe, Bedding, Casagrande, Handberg,
  Lund, Nissen, Chaplin, Huber, Serenelli, Stello, Van~Eylen, Campante,
  Elsworth, Gilliland, Hekker, Karoff, Kawaler, Kjeldsen, \&
  Lundkvist}]{silvaaguirre2015}
Silva~Aguirre, V., Davies, G.~R., Basu, S., {et~al.} 2015, Monthly Notices of
  the Royal Astronomical Society, 452, 2127, \dodoi{10.1093/mnras/stv1388}

\bibitem[{Simonian {et~al.}(2019)Simonian, Pinsonneault, \&
  Terndrup}]{simonian2019}
Simonian, G. V.~A., Pinsonneault, M.~H., \& Terndrup, D.~M. 2019, The
  Astrophysical Journal, 871, 174, \dodoi{10.3847/1538-4357/aaf97c}

\bibitem[{Simonian {et~al.}(2020)Simonian, Pinsonneault, Terndrup, \&
  Van~Saders}]{simonian2020}
Simonian, G. V.~A., Pinsonneault, M.~H., Terndrup, D.~M., \& Van~Saders, J.~L.
  2020, The Astrophysical Journal, 898, 76, \dodoi{10.3847/1538-4357/ab9a43}

\bibitem[{Skumanich(1972)}]{skumanich1972}
Skumanich, A. 1972, The Astrophysical Journal, 565

\bibitem[{Somers {et~al.}(2017)Somers, Stauffer, Rebull, Cody, \&
  Pinsonneault}]{somers2017}
Somers, G., Stauffer, J., Rebull, L., Cody, A.~M., \& Pinsonneault, M.~H. 2017,
  The Astrophysical Journal, 850, 134, \dodoi{10.3847/1538-4357/aa93ed}

\bibitem[{Spada \& Lanzafame(2020)}]{spada2020}
Spada, F., \& Lanzafame, A.~C. 2020, Astronomy \& Astrophysics, 636, A76,
  \dodoi{10.1051/0004-6361/201936384}

\bibitem[{Tayar {et~al.}(2020)Tayar, Claytor, Huber, \& {van
  Saders}}]{tayar2020}
Tayar, J., Claytor, Z.~R., Huber, D., \& {van Saders}, J. 2020,
  arXiv:2012.07957 [astro-ph].
\newblock \doarXiv{2012.07957}

\bibitem[{{Thoul} {et~al.}(1994){Thoul}, {Bahcall}, \& {Loeb}}]{Thoul1994}
{Thoul}, A.~A., {Bahcall}, J.~N., \& {Loeb}, A. 1994, \apj, 421, 828,
  \dodoi{10.1086/173695}

\bibitem[{Townsend \& Teitler(2013)}]{townsend2013}
Townsend, R. H.~D., \& Teitler, S.~A. 2013, Monthly Notices of the Royal
  Astronomical Society, 435, 3406, \dodoi{10.1093/mnras/stt1533}

\bibitem[{{van Saders} {et~al.}(2016){van Saders}, Ceillier, Metcalfe, Aguirre,
  Pinsonneault, Garc{\'i}a, Mathur, \& Davies}]{vansaders2016}
{van Saders}, J., Ceillier, T., Metcalfe, T.~S., {et~al.} 2016, Nature, 529,
  181, \dodoi{10.1038/nature16168}

\bibitem[{van Saders \& Pinsonneault(2013)}]{vansaders2013}
van Saders, J., \& Pinsonneault, M.~H. 2013, The Astrophysical Journal, 776,
  67, \dodoi{10.1088/0004-637X/776/2/67}

\bibitem[{{van Saders} {et~al.}(2019){van Saders}, Pinsonneault, \&
  Barbieri}]{vansaders2019}
{van Saders}, J., Pinsonneault, M.~H., \& Barbieri, M. 2019, The Astrophysical
  Journal, 872, 128, \dodoi{10.3847/1538-4357/aafafe}

\bibitem[{{Vitense}(1953)}]{vitense1953}
{Vitense}, E. 1953, \zap, 32, 135

\bibitem[{Weber \& Davis(1967)}]{weber}
Weber, E.~J., \& Davis, L. 1967, The Astrophysical Journal, 148

\bibitem[{White {et~al.}(2017)White, Benomar, Silva~Aguirre, Ball, Bedding,
  Chaplin, {Christensen-Dalsgaard}, Garcia, Gizon, Stello, Aigrain, Antia,
  Appourchaux, Bazot, Campante, Creevey, Davies, Elsworth, Gaulme, Handberg,
  Hekker, Houdek, Howe, Huber, Karoff, Marques, Mathur, McQuillan, Metcalfe,
  Mosser, Nielsen, R{\'e}gulo, Salabert, \& Stahn}]{white2017}
White, T.~R., Benomar, O., Silva~Aguirre, V., {et~al.} 2017, Astronomy \&
  Astrophysics, 601, A82, \dodoi{10.1051/0004-6361/201628706}

\bibitem[{Woitaszek {et~al.}(2009)Woitaszek, Metcalfe, \&
  Shorrock}]{Woitaszek2009}
Woitaszek, M., Metcalfe, T., \& Shorrock, I. 2009, Proceedings of the 5th Grid
  Computing Environments Workshop on - GCE '09, 1,
  \dodoi{10.1145/1658260.1658262}

\end{thebibliography}
\bibliographystyle{aasjournal}

\appendix

\section{Neural Network Structure} \label{sec:nn_structure}

As described in \S \ref{sec:ANN}, our artificial neural network was generated with six hidden layers of 128 neurons. A model summary can be found in Table \ref{tab:nn_summary}.

\begin{table}[ht!] 
\footnotesize
\centering
\begin{tabular}{lll} 
\toprule
Layer (type)                   & Output Shape                & $N_{\rm params}$    \\ \midrule 
normalization (Normalization)  & (None, 7)                   & 15         \\ 
dense (Dense)                  & (None, 128)                 & 1024       \\ 
dense\_1 (Dense)                & (None, 128)                 & 16512      \\ 
dense\_2 (Dense)                & (None, 128)                 & 16512      \\ 
dense\_3 (Dense)                & (None, 128)                 & 16512      \\
dense\_4 (Dense)                & (None, 128)                 & 16512      \\ 
dense\_5 (Dense)                & (None, 128)                 & 16512      \\ 
dense\_6 (Dense)                & (None, 4)                   & 516        \\ 
rescaling (Rescaling)          & (None, 4)                   & 0          \\ \midrule
Total params: 84,115 \\ 
Trainable params: 84,100 \\ 
Non-trainable params: 15 \\ 
\bottomrule
\end{tabular} 
\caption{Model summary for our ANN.} 
\label{tab:nn_summary} 
\end{table}

\section{Model Validation} \label{sec:validation}

\begin{figure*}[ht!]
    \centering
    \includegraphics[width=.3\textwidth]{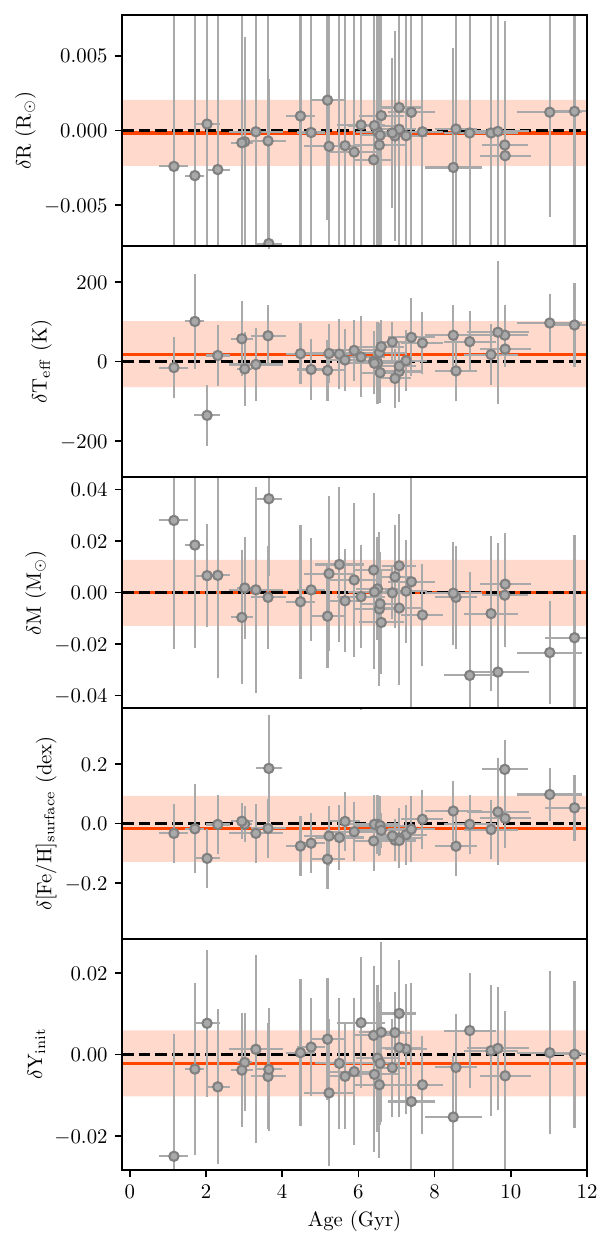}
    \caption{Difference between observed values of stellar properties and predictions from our probabilistic model (predicted$-$observed), plotted against stellar age. The red line shows the median of the difference, with the standard deviation shown by the shaded region.}
    \label{fig:params}
\end{figure*}

To validate the performance of our model, we calculated the difference between the observed value and the median of the posterior sampled distribution for each parameter in our grid. Figure \ref{fig:params} shows this value, where $\delta X = X_{\rm predicted} - X_{\rm observed}$ for a parameter $X$. We find good overall agreement between predicted and observed values, with no significant systematic offsets. 

\begin{figure*}[ht!]
    \centering
    \includegraphics[width=.75\textwidth]{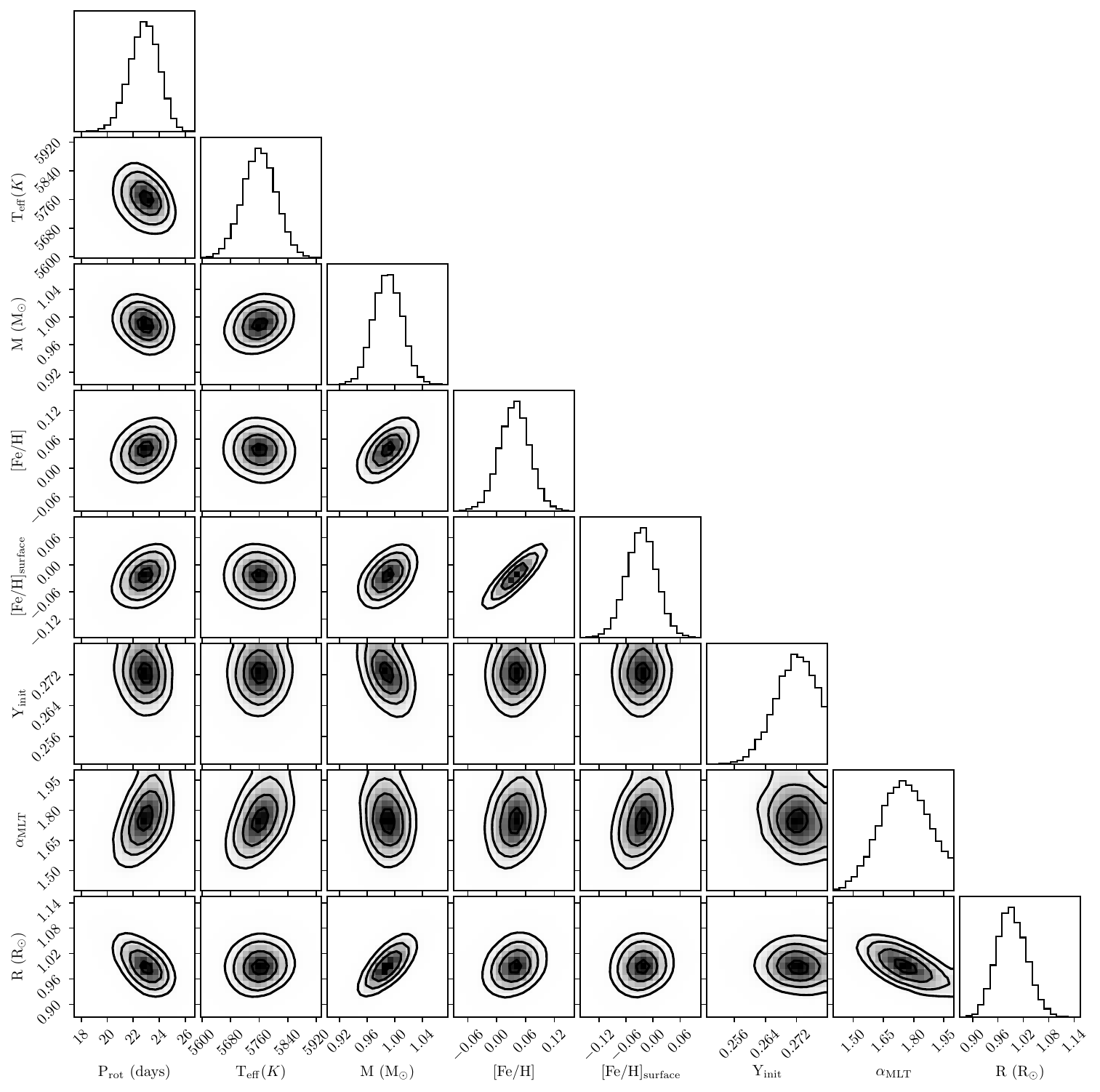}
    \caption{Corner plot showing marginal and joint posterior distributions for the Sun from our sampled probabilistic model.}
    \label{fig:sun_corner}
\end{figure*}

For each star in our sample, we retrieve full posterior distributions for each parameter. In Figure \ref{fig:sun_corner}, we show the sampled marginal and joint posterior distributions for the Sun.

\end{document}